\documentclass[letterpaper,twocolumn,10pt, top=1.5cm, bottom=2cm, left=1.5cm, right=1.5cm]{article}
\usepackage{usenix-2020-09}
\usepackage{titling}

\usepackage{graphicx}
\usepackage{multirow}
\usepackage{tikz}
\usepackage{amsmath}
\usepackage{booktabs}
\usepackage{xspace}
\usepackage{verbatim} 
\usepackage{appendix}
\usepackage{booktabs} 
\usepackage{tabularx}
\usepackage{amssymb}
\usepackage{caption}
\usepackage{subcaption}
\usepackage{balance}
\graphicspath{{./images/}}

\newcommand{\pname}{{\texttt{EnCortex}}\xspace}
\newcommand{\databackend}{{\texttt{DataBackend}}\xspace}
\newcommand{\dfbackend}{{\texttt{DFBackend}}\xspace}
\newcommand{\dbbackend}{{\texttt{DBBackend}}\xspace}
\newcommand{\repo}{Code available at \href{https://anonymous.4open.science/r/encortex/}{https://anonymous.4open.science/r/encortex/}}

\usepackage[inline]{enumitem}
\newcommand{\rom}[1]{(\lowercase\expandafter{\romannumeral #1\relax})}
\newcommand{\ROM}[1]{(\lowercase\expandafter{\romannumeral #1\relax})}
\newcommand{\xmark}{\ding{55}}%

\usepackage{pifont}
\DeclareRobustCommand{\circled}[1]{\ding{\the\numexpr #1 + 171 \relax}}
\DeclareRobustCommand{\circleddark}[1]{\ding{\the\numexpr #1 + 201 \relax}}

\usepackage{filecontents}
\usepackage{hyperref}
\hypersetup{breaklinks=true}

\begin{document}




 \author{
 {\rm Millend Roy\textsuperscript{*$\dagger$}}, 
 {\rm Vaibhav Balloli\textsuperscript{*$\dagger$}}, 
 {\rm Anupam Sobti\textsuperscript{$\dagger$}},
 {\rm Srinivasan Iyengar\textsuperscript{$\ddagger$}},\\
 {\rm Shivkumar Kalyanraman\textsuperscript{$\ddagger$}},
 {\rm Tanuja Ganu\textsuperscript{$\dagger$}},
 {\rm Akshay Nambi\textsuperscript{$\dagger$}}\\
 \textsuperscript{$\dagger$}Microsoft Research, \textsuperscript{$\ddagger$}Microsoft \\
} 




\setlength{\droptitle}{-1.5cm} 
\title{\Large \bf EnCortex: A General, Extensible and Scalable Framework for Decision Management in New-age Energy Systems}
\date{}
\maketitle
\renewcommand{\thefootnote}{\fnsymbol{footnote}}
\footnotetext[1]{Authors have equal contribution.}
\renewcommand{\thefootnote}{\arabic{footnote}}
\begin{abstract}
With increased global warming, there has been a significant emphasis to replace fossil fuel-dependent energy sources with clean, renewable sources. These new-age energy systems are becoming more complex with an increasing proportion of renewable energy sources (like solar and wind), energy storage systems (like batteries), and demand side control in the mix. Most new-age sources being highly dependent on weather and climate conditions bring about high variability and uncertainty. Energy operators rely on such uncertain data to make different planning and operations decisions periodically, and sometimes in real-time, to maintain the grid stability and optimize their objectives (cost savings, carbon footprint, etc.). Hitherto, operators mostly rely on domain knowledge, heuristics, or solve point problems to take decisions. These approaches fall short because of their specific assumptions and limitations. Further, there is a lack of a unified framework for both research and production environments at scale.

In this paper, we propose \pname to address these challenges. \pname provides a general, easy-to-use, extensible, and scalable energy decision framework that enables operators to plan, build and execute their real-world scenarios efficiently. We show that using \pname, we can define and compose complex new-age scenarios, owing to industry-standard abstractions of energy entities and the modularity of the framework. \pname provides a foundational structure to support several state-of-the-art optimizers with minimal effort. \pname supports both quick developments for research prototypes and scaling the solutions to production environments. We demonstrate the utility of \pname with three complex new-age real-world scenarios and show that significant cost and carbon footprint savings can be achieved.
\end{abstract}

\section{Introduction}
\label{sec:intro}

With increasing global climate change, the energy industry is going through a paradigm shift~\cite{bergmann2006valuing,rolnick2022tackling}. Electricity generation is currently responsible for 40\% of global CO2 emissions~\cite{doerr2021speed}. Hence, there is an increasing focus on incorporating greener and renewable energy sources like solar, wind, and hydro in the energy generation mix. The share of renewables in global electricity generation jumped to 29\% in 2020 and is expected to grow up to 69\% by 2050~\cite{doerr2021speed}, \cite{renewables2021iea}. Being dependent on the weather conditions like wind, cloud covers, and rains, these renewable energy sources induce high variability and intermittency in the generation mix~\cite{bremen2010large}. To address such challenges, various types of energy storage systems (ESS) such as batteries, hydroelectric pumps, and flywheels are also introduced~\cite{ruhnau2022storage}. 
On the other side, energy demand still has an increasing trend with a 19\% increase in energy demand between 2009 to 2019{~\cite{renew_data}}. There is also significant uncertainty in the energy demand depending upon seasonality, economic trends, weekday/weekend, etc. There are also various types of demand-side management (DSM) approaches (including various control actions to reduce or increase energy demands) that are being executed by consumers or energy operators to reduce the peak demands and increase monetary savings~\cite{eiademandside}.

\begin{figure}
    \centering
    \includegraphics[scale=0.5]{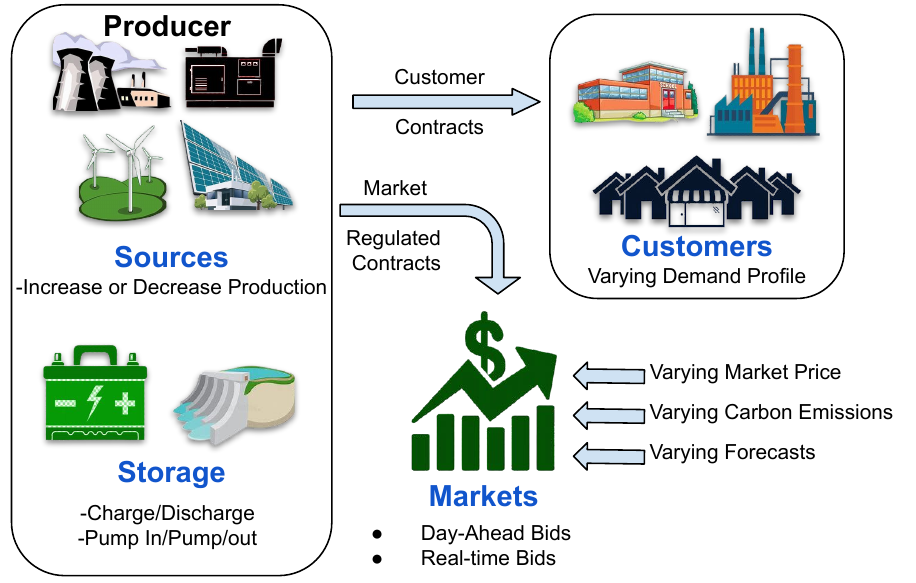}
    \caption{Illustration of an energy producer with the complexity involved in the decision-making.}
    \label{fig:problem-illustration}
\end{figure}

Thus, new-age energy systems have become more complex with increasing uncertainty from the generation sources and demand sides. Various energy operators -- like energy producers (i.e., solar/ wind farm operators), balancing authorities or consumers like (data centers, malls, universities, or households) need to make different planning and operations decisions periodically, and sometimes in near real-time, to maintain the grid stability and to optimize their objectives (maximize cost savings or reduce carbon footprint). Also, to support such a paradigm shift, new types of green energy markets and incentives are being introduced making the whole system dynamic and adapting to the changes rapidly.

Improving the efficiency of operating power plants has been a long-studied problem~\cite{vpp-survey} in the academic literature. There are various optimization scenarios studied in the literature, including participation in multiple markets~\cite{multiple-markets-bidding}, demand response and surplus management~\cite{day-ahead-bidding}, as well as other scenarios with multi-objective optimization for minimization of carbon emissions as well as monetization~\cite{joint-emission-monetization-framework}. Various techniques of optimization, e.g., linear programming, robust optimization, reinforcement learning, fuzzy logic, etc. have been used to address the optimization problem~\cite{vpp-survey}. However,  industries and energy operators still rely on heuristics and/or manual methods or solving point problems for making their planning and operations decisions.

A primary reason for this gap stems from the lack of generalization, customizability, and ease of use with the existing approaches{~\cite{gap1,vpp-survey}}. Figure~\ref{fig:problem-illustration} shows a producer with a portfolio of generation sources like wind, solar, and storage. Different types of sources have different actions available, e.g., charge/discharge for batteries, increase/decrease production for nuclear/thermal plants, or, just curtail output for solar/wind farms. Producers also participate in various markets - long-term purchase agreements, week-ahead, day-ahead, intra-day and real-time markets~\cite{burger2014managing}. The producer might also have contracts with different classes of consumers - data centers, residential, commercial, industrial, etc{~\cite{source1}}. Each of these contracts has its own penalty structures and deadlines. In this context, the producer has to make decisions corresponding to the available actions while incorporating the varying generation profiles, varying consumer profiles, and the fluctuation in markets over a long-term as well as the short-term bidding process. The uncertainty in prices, demand, and generation makes this problem even harder to address without a generalized optimization framework. Further, the optimization decision can vary significantly when the scenario being optimized has an additional power plant or a long-term contract with variable demand throughout the day. This makes it difficult for existing formulations to be useful for a new scenario and end up being bespoke solutions that aren't applicable. 
Therefore, there is a need for a general framework that provides - (i) \textit{industry-standard abstractions} for various types of sources (modeling their actions and behaviors in a consistent way), (ii) \textit{extensibility support} to introduce new types of energy entities (like hydrogen electrolyzer), (iii) \textit{composibility} for developing new scenarios based on energy operators' needs, constraints and objectives, (iv) inherent ability to make \textit{data-driven decisions and perform what-if analysis}, and most importantly, (v) \textit{ease of use} for developers, business users, and energy operators. 

To address the above challenges, we design \pname, a general, extensible, composable, and scalable framework for decision management in sustainable planning and operations of new-age energy systems. \pname provides industry-standard abstractions with realistic modeling and configurations for most new-age entities. \pname supports both the definition and composition of new scenarios in a structured way with very minimal overhead. \pname offers native support to model the variability and uncertainty in renewable sources and markets by incorporating data-centric decision-making. Finally, \pname is easy to use, programmable, and scalable to run complex scenarios across multiple compute clusters and provides support for different optimizers to perform what-if analysis. 

We summarize our contributions as follows: 
\circleddark{1} \pname is the first general, easy-to-use, extensible, and scalable energy decision framework that enables various energy operators to plan, build and execute their real-world energy decision scenarios efficiently. Energy operators can quickly express their objectives and constraints without worrying about the details of implementation and scaling (Section~\ref{sec:overview}). 
\circleddark{2} \pname is modular and data-centric by design with support for public datasets and streaming data 
(Section~\ref{sec:infra_components}). 
\circleddark{3} We demonstrate the support for modularity, extensibility, and composability features in \pname, by formulating three popular new-age energy scenarios with different levels of complexity (Section~\ref{sec:scenario_overview}).
\circleddark{4} We show that \pname's core architecture provides a foundational structure and support for both traditional and neural-network-based optimizers. Our experiments demonstrate significant savings in both cost and carbon footprint across datasets/scenarios (Section~\ref{sec:results}). Finally, our experiments illustrate scenarios with industry partners in various geographies around the world. Thus giving confidence in the generality and applicability of \pname to a large number of practical new-age energy scenarios.

\section{Motivation}
\label{sec:motivation}
\pname provides a foundational structure for generic scenario compositions as opposed to point formulation with support for ease of use, extensibility, composability, and scalability. In this section, we describe why such a system is needed to aid energy operators. 
\subsection{A Motivating Example}
\label{sec:motivation_eg}




Most renewable energy resources such as solar and wind, being highly dependent on weather and climate conditions, bring about variability and uncertainty~\cite{bremen2010large}. Hence, this has led to the rise of Energy Storage Systems (ESS), which provide a multitude of low-carbon emission-based ancillary services to the energy operators such as demand peak-shaving and grid stabilization~\cite{bss}. 

Our partner, a global energy provider is deploying Li-ion batteries at their consumer premises to either maximize economic profit (increase cost savings by charging/discharging the battery at off-peak/peak price periods) or maximize environmental impact (reduce overall carbon footprint by charging/discharging the battery based on usage of renewable/non-renewable energy sources). Thus, it is important to optimize the management and scheduling of these batteries (when to charge or discharge) to maximize the objective. Intelligent optimization techniques are required to recommend optimal schedules that take into account numerous real-world factors such as available generation mix, state of charge of batteries, their efficiency, electricity purchase, and sale contracts, etc. 

In order to determine the optimal schedule of the deployed batteries, our partner has to build this solution from scratch as prior solutions are mostly customer-specific and are not generalizable or reusable{~\cite{inti,vpp-survey}}. Specifically, our partner has to define the different entities involved in the scenario such as the battery, grid, and renewable generation mix along with defining the objective function (price or carbon savings) and build an optimization service using either standard robust optimization techniques or newer neural network-based approaches. Developing an end-to-end solution for this scenario is non-trivial and 
 majority of energy operators find it significantly hard, and time-consuming.  

\textbf{Problem 1: Several point solutions exist, but fall short because of their specific assumptions and limitations.}
While the above scenario is pretty common and representative, there is no off-the-shelf tool/framework to solve this. Energy operators resort to developing ad-hoc tools to handle specific cases leading to inefficient and complex implementations. These bespoke solutions are mostly developed with specific assumptions on the abstractions and definitions of the sources and battery, and optimization algorithms employed. For instance, it is non-trivial to change or add new storage definitions say from Li-ion to another battery composition.

\textbf{Problem 2: The ad-hoc solutions are inefficient and usually cannot be reused.}
It is clear that the above point/ad-hoc solutions are not built to be applied to other configurations. The scenario definitions, energy purchase and sell contracts, and entity abstractions will likely be specific for that formulation. Since the design and implementation have to be swift, reusability, performance and scalability will likely be not the priorities. {Thus, requiring significant understanding and time to re-purpose these solutions.}  

\textbf{Problem 3: Existing solutions are not modular and do not support the addition of custom scenario definitions. }
Even if there exists a tool/solution to solve the above scenario, modifying the scenario objective and adding new entities/markets is still a huge challenge. For example, in the above scenario let's say the partner wants to develop a multi-objective function to maximize both cost and carbon footprint instead of just maximizing cost or carbon footprint. In such cases given the lack of modularity, abstractions, and extensibility of ad-hoc solutions it is impossible to compose and define such custom scenarios. 

After observing these problems with our partners during planning and operations, we believe that a general, easy-to-use, extensible, and data-centric decision management framework can bring significant benefits to energy operators. It can help them construct, understand, and analyze different scenarios swiftly, and make decisions more efficiently.   
\subsection{Design Goals}
\label{sec:design_goals}
Motivated by many real-world examples like the one in~Section~\ref{sec:motivation_eg}~(and more in~Section~\ref{sec:scenario_overview}), we derive the design goals of \pname to support numerous new-age energy scenarios.

\textbf{1. New-age industry-standard entity abstractions.} Hitherto, most of the solutions employ unstructured and disjoint definitions for energy entities such as sources, storage, consumer profiles, and markets~\cite{batteries_grid}. These definitions are not standardized and cannot be reused for quick development. A key design goal is to build industry-standard abstractions that incorporate how the new-age entities such as sources, storage, and markets can be modeled realistically. 

\textbf{2. Extensibility support for entity abstractions.} While entity abstractions are useful, it is even more important to support the extensibility of these abstractions. For example, a market abstraction should support extensions to address different types of markets such as intra-day, day-ahead, and real-time. Thus allowing extensions for the well-defined generic abstractions and support for definitions of new entities. 

\textbf{3. Composability of new-age energy scenarios.} A scenario is comprised of numerous entities. For example, in the energy arbitrage scenario (Section~\ref{sec:motivation_eg}), the entities include a battery, generation sources (solar, wind), and electricity purchase contracts. Currently, scenario formulation is not standardized and extensible. A key design goal is to support easy-to-use and intuitive ways to compose a scenario with entity abstractions. Further, we should be able to alter/modify these scenario definitions based on the application needs. For example, it should be trivial to extend single objective optimization (such as carbon footprint reduction) to  multi-objective (i.e., joint optimization of both cost and carbon footprint).  

\textbf{4. \label{itm:modular-custom} Modular and support for custom scenario definitions.} Current solutions are mostly bespoke and cumbersome to expand{~\cite{inti}}. A general framework should be capable of handling a multitude of scenario compositions with different entities, objectives, and constraints. 

\textbf{5. \label{itm:data-centric} Data-centric decision-making with support for real-time analysis.}
Data-based decision-making is a key element to optimal decision-making. Existing solutions either employ simulated data or work with small datasets, which may not capture the complete trend, variability, and uncertainties. Hence, framework design should provide native support to ingest large entity datasets of solar, wind, price information, etc., to perform detailed analysis and derive optimal decisions at different time granularity.  

\textbf{6. \label{itm:scalable-workflow} Scalable workflow to aid what-if analysis.}
Given the lack of modularity in the existing optimization solutions, they typically do not scale well, especially when evaluating on large datasets. Hence, it is important to provide a host of tools in the framework to support scalable workflows, offload compute to CPU/GPU clusters, and native support for different optimizers to do detailed what-if analysis. 

\textbf{7. \label{itm:ease-of-use} Ease of use.} 
Energy operators might have limited data science expertise and hence it is important for the framework to be easy to use wrt defining and composing scenarios, parameterizing and adding new abstractions, and supporting quick development for research prototypes to even scaling in production environments. Further, it should support visualizing the scenario and interactive result generation to understand the decision outcomes on different datasets and time periods. 
\section{\pname Overview}
\label{sec:overview}
\pname is an end-to-end framework for sustainable operations and planning in new-age energy systems. At a high level, {to} support \texttt{operations}, live data from entities such as solar, wind farms, consumer demands are connected to the framework and the corresponding actions for optimal control with respect to a specific objective, e.g., profit maximization are sent back to the entities, e.g., battery actions, curtailment actions for solar/wind, etc. These are then handled by the associated SCADA systems of these entities and executed on the appropriate hardware. For \texttt{planning} tasks, arbitrary entities like additional customers, and energy generating plants (solar, wind, hydro, etc.) can be added to the scenario and the effect on revenues/penalties can be estimated. We provide various functionalities to support realistic implementation and detailed what-if analysis for such planning scenarios. We now describe the components and the architecture of \pname. 
\subsection{Core components}
\label{sec:corecomponents}
\pname has three core components, \textit{viz.,} entities, contracts, and decision units. 
\subsubsection{Entities}
\label{sec:entities}
\begin{figure}
	\centering
	\begin{minipage}[t]{0.2\textwidth}
		\centering
		\includegraphics[width=0.95\textwidth, height=1.4in]{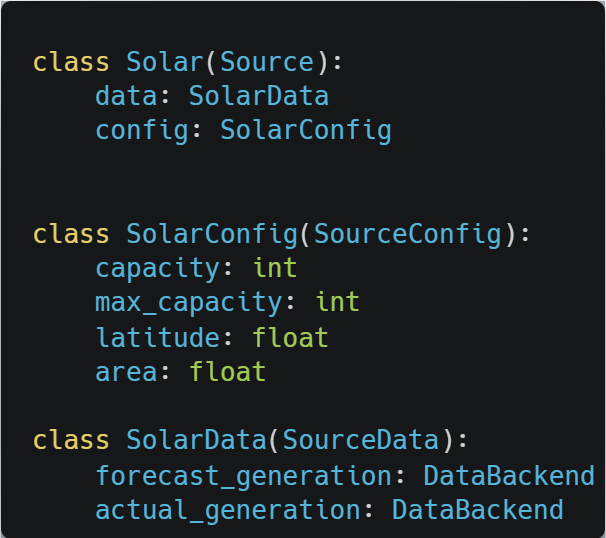}
		\subcaption{Solar entity}
		\label{fig:solar}
	\end{minipage}%
	\hspace{1ex}
\begin{minipage}[t]{0.2\textwidth}
		\centering
		\includegraphics[width=0.95\textwidth]{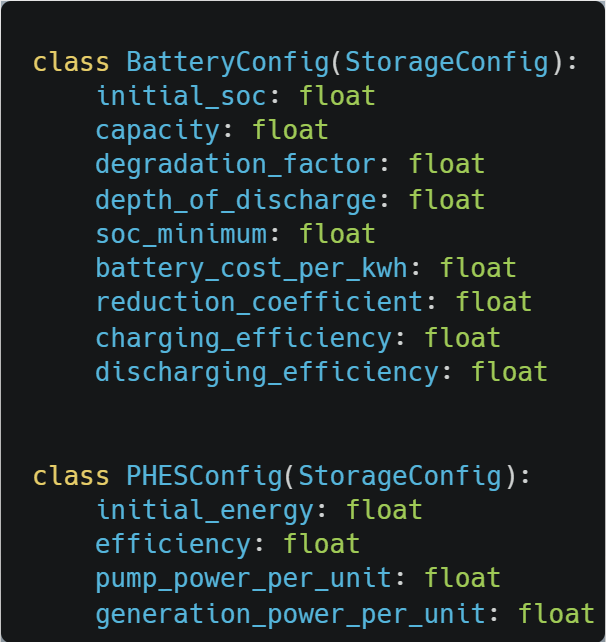}
		\subcaption{Battery entity}
		\label{fig:battery}
	\end{minipage}
		\caption{Example abstractions defined in \pname.}
		\label{fig:abs_eg}
\end{figure}

Entities are the backbone of our framework. An entity could be any component that consumes/generates electricity. At a high level, each entity has four key attributes: (i) Data: It is a data structure containing various data fields related to the component, (ii) Actions:  defines how to  implement supported actions associated with the component, (iii) Configuration: to support config-based initialization and (iv) Schedule: to define time-based deadlines for each component. Entities are:\\
\circleddark{1} \textbf{\texttt{Sources}:} A source is any component that is capable of generating electricity. Solar, Wind, Thermal, Nuclear, Hydro, etc., are all examples of the source class. Sources have \textit{forecast\_generation} and optionally \textit{actual\_generation} attributes {as a part of the data field}. In some cases, actions might be associated with sources. Figure~\ref{fig:abs_eg} shows an example and Appendix~\ref{appendix:abstract} has additional details.\\
\circleddark{2} \textbf{\texttt{Consumers}}: A consumer is a component that consumes energy, e.g., a household, campus, shopping mall, etc. Similar to sources, consumers are a majorly data-only entity, but actions can also be associated with a consumer in case of demand response. \\
\circleddark{3} \textbf{\texttt{Markets}}: A \textit{market} is an entity where energy can be sold or purchased. Typically, this is done through a bidding/continuous auction process{, which defines the action space of a market entity i.e. simultaneous price bidding and volume allocation}. Different markets across the world serve different purposes, e.g., a day-ahead market is used to provide an estimate of energy availability one day ahead of time. A detailed discussion of the types of markets can be found here~\cite{vpp-survey}. 
To model all these types of markets, we simplify the behavior down to a schedule definition. Corresponding to each market, there are periods of time where the bidding is open. The end of this window is the deadline before which certain values have to be committed to a market. We use a cron-style definition of schedule to define the number of variables that have to be predicted before the deadline. {See Appendix~\ref{appendix:market_def} for an example of the schedule definition.}\\
\circleddark{4} \textbf{\texttt{Storage}}: The storage class is inherited from a source class. The only exception here is that negative generation values are allowed. The configuration includes factors related to charging rates, degradation rates and capacities. Battery Storage Systems (BSS), Pumped Hydro Systems (PHS) are all modeled as storage entities (see Figure~\ref{fig:abs_eg}) thus showcasing the extensibility feature of the provided abstractions. \\
\circleddark{5} \textbf{\texttt{Producer}}:
A producer is associated with multiple entities which have contracts with various consumers and markets.

These entities can be either, \textit{offline} (has representative data in file/database), \textit{online} (capture streaming data from end devices) or \textit{simulated} (physics-based models to generate data). 

\subsubsection{Contracts}
A \textit{contract} is required to define the flow of energy between two entities in the framework. A source supplying energy to a consumer should have a contract with the fields - \textit{min\_supply}, \textit{max\_supply}, and a penalty function. A penalty function is a function of supply and demand over a time horizon. For example, a linear penalty is applied with the deviation in the promised supply at every instant, i.e., let's say for every 15 minutes, if the demand is more than the supply, a fixed cost of $P\$/kWh$ is levied on the source.
The penalty function is a combination of the penalty values over several time horizons.

\subsubsection{Decision Units}
\label{sec:decision_unit}
Based on the defined entities and contracts associated with a particular producer, we identify subsets of entities and contracts where the decision-making is dependent on each other. We use a graph representation of entities as nodes and contracts as edges to identify decision units. A decision unit generates critical information on the schedule and the associated actions based on the included contracts/entities.


In Section~\ref{sec:scenario_overview} we show how these components are used to define and compose different new-age scenarios.

\begin{figure}[!t]
\centering
\includegraphics[width=0.49\textwidth, height =1.8in]{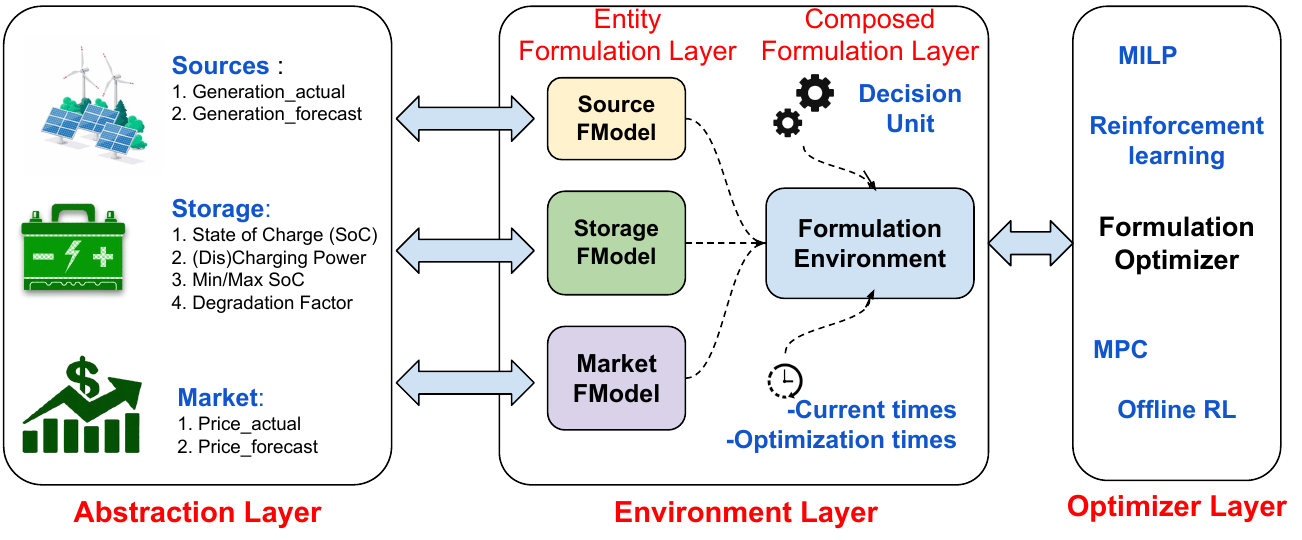}
\caption{Overview of \pname architecture.}
\label{fig:env-architecture}
\end{figure}
\subsection{\pname Architecture}
\label{sec:architecture}
Our framework has three layers (Figure~\ref{fig:env-architecture}), which are used for composing new-age scenarios of different complexities:

\textbf{1. Abstraction Layer}: An abstraction layer uses entity-specific configurations to define entity behavior required for a scenario. It uses the framework provided unified data handling capabilities (Section \ref{sec:infra_components}) to handle data from all types of underlying objects, be it file-based, a web endpoint, or custom instrumentation in a live plant. 
Essentially, the abstraction layer takes care of all the data associated with a particular entity {and simulation of recommended action} which is based on the physics of the entity. 

\textbf{2. Environment Layer}: {The key purpose of this layer is to provide data and state information from entities(called state space) that are needed to make a decision and a central point to orchestrate all the required decisions based on the schedule (termed as action space).}
The environment layer has an entity-specific component called the \textit{Entity Formulation Layer} which exposes the micro variables like volume output, emission output, etc. in a uniform API to the environment. The decision unit identified (Section \ref{sec:decision_unit}) is used as an input along with the entity formulation layer and a user-specified optimization horizon to generate the formulation environment. This layer identifies the time period for which models have to be built, maintains the current time in simulation and incorporates all the variables and actions that are required to solve the optimization for this particular time period. This environment composes the scenario based on the decision unit under consideration (called the \textit{Composed Formulation Layer}) to expose the macro variables like revenue, penalty, transmission penalty, emissions, etc. {When an action is executed in simulation, {the agents} receive a reward indicating how they've performed on a singular instance of the objective function that needs to be maximized.} Appendix~\ref{appendix:env_layer} includes details of entity formulation and composed formulation layer.

\textbf{3. Optimizer Layer}: An optimizer layer finally uses the environment layer outputs (revenue, penalty, transmission penalty, emissions, etc.) for every timestamp and runs the optimization based on the objective defined (e.g., monetization). \pname provides native support for various state-of-the-art algorithms apart from bringing your own optimizer code. The energy operator can run numerous experiments to determine which algorithm provides the best optimal schedules while maximizing the objectives. The supported algorithms include Simulated Annealing (SA), Mixed Integer Linear Programming (MILP), and Deep Q Network-based Reinforcement Learning (DQN-RL) (see Appendix~\ref{appendix:opt} for details). \textit{In this work, we have developed adaptations to the above algorithms to be compatible for new-age energy scenarios (Section~\ref{sec:scenario_overview}).}

\section{\pname Infrastructure Components}
\label{sec:infra_components}
\pname is developed as a modular, easy-to-use, and ready-to-deploy framework. We now describe the infrastructure components developed to support these functionalities.  

\textbf{{{Storage Backends:}}} The storage backend module helps keep a centralized data API to support data operations across various abstractions, entities, and scenarios. \pname supports two storage backends that inherit \databackend - an abstract class to deal with various \texttt{CRUD} operations, \textit{viz.,} {\textit{(i) \dfbackend:}} This is built on top of \texttt{pandas}'s \texttt{DataFrame} and \texttt{Series} API to support offline, tabular datasets. {\textit{(ii) \dbbackend:}} The \dbbackend helps connect to various relational databases - SQLite, MySQL, etc., to access data. 

\textbf{{{Pipelines:}}}
Each energy operator might have their own set of operational and deployment constraints. Depending on the data availability, frequency of incoming data, SLA requirements, and choice of optimizers, orchestrating pipelines becomes crucial. In \pname we have developed \texttt{Runners} to help in orchestrating time-based decisions and are designed to run independently on separate threads on the same machine or on different machines altogether. Broadly there are two types of runners, \textit{(i) Scheduled Runner} enables running optimization on a fixed schedule, as a co-routine. Scheduled runners are best suited when data is \textit{assumed} to be available or when SLAs must be strictly adhered to. {\textit{(ii) Event-based Runner:} In scenarios where data is received from a third-party, an event-based runner helps to run optimizations that are triggered when new data is received. }


\textbf{{{DataLoaders:}}}
DataLoaders provides a uniform API to load datasets available publicly or versioned datasets stored on the cloud. The dataloaders are built on top of storage backends to have consistent data handling across the framework. These data loaders are pythonic and support both iterators-based data loading for batch-based operations as well as a query-like interface for sequential data-loading. 
We also add cloud-based support to load datasets through various frameworks like Ray~\cite{moritz2018ray} and Dask~\cite{rocklin2015dask}.


\textbf{{Forecasting Models:}} Decisions within the framework are based on data received from forecasting models/services. We have developed a unified forecasting module, that supports numerous state-of-the-art time-series forecasting techniques and also operators can bring their own forecaster model and integrate it. We support a wide range of techniques from simple heuristics to state-of-the-art algorithms, i.e., 
{\textit{(i) Noise forecast:} gaussian noise is added to actual data} ,\textit{ (ii) Yesterday's forecast:} previous day's actual data as forecast for the next day,\textit{ (iii) Mean:} mean of actual data for previous N days as forecast, \textit{(iv) LightGBM\cite{ke2017lightgbm}}, a decision tree based algorithm that builds a tree-like structure to perform regression and \textit{(v) N-BEATS\cite{oreshkin2019n}}, a purely neural network-based model.


\textbf{{{EnCortex Configuration and CLI:}}}
To instantiate the abstractions supported by \pname, we have built a command-line utility (CLI) that takes configuration files as input. We utilize Hydra~\cite{Yadan2019Hydra}, which allows us to modularize configurations of different entities allowing us to define a config-based execution that can be overridden using configuration files.


\textbf{{{Visualization:}}}
Visualizing and tracking various performance metrics and decisions helps showcase the decisions and capabilities of various agents to energy operators. Therefore, we provide a command-line interface to launch a browser-based dashboard to understand scenario composition, plot, and track various optimizers' performance and analyze the rewards/penalties on different datasets and time periods. 

\textbf{{{Machine Learning Operation:}}}
When deploying neural network-based optimizers, model deployment, reproducible pipelines, and other principles of Machine Learning Operations (MLOps) for efficient workflows are critical. We have built-in support for different MLOps such as model and dataset versioning, experiment tracking, and reusable pipelines. 

\textbf{{{Callbacks:}}}
Callbacks are designed in \pname to provide an additional level of control and modify various aspects of the pipeline. The key callbacks are: \textit{(i) Logging and Monitoring callbacks} captures detailed logs to efficiently monitor and debug the pipelines. \textit{(ii) Closed loop callbacks} allow energy operators to connect the decision/action derived by the optimizers with physical controllers to execute the decisions made. \textit{(iii) Connect to other applications callback} facilitates other applications/scenarios to easily build on top of \pname. \textit{(iv) Safety handling callbacks} provide easy-to-use callbacks that are triggered during potential safety warnings to help energy operators to analyze and debug the issues quickly. 


\section{Real-world New-age Energy Scenarios}
\label{sec:scenario_overview}
We now present a few prevalent new-age energy scenarios based on discussions with our global industry partners. We show how \pname framework is employed to compose, analyze and derive optimal decisions. 
\subsection{Scenario 1 - Energy Arbitrage (EA)}
\label{sec:energy_arbitrage}
The variability and uncertainty associated with renewable sources can cause reliability issues in the energy grid. Thus new-age scenarios now include an energy storage system to assist in supply and demand mismatch. Energy Arbitrage (EA) can be simply defined as purchasing more electricity during Off-peak periods, storing that electricity via energy storage systems, and discharging it during Peak periods. 

\begin{figure}
    \centering
    \includegraphics[width=\columnwidth]{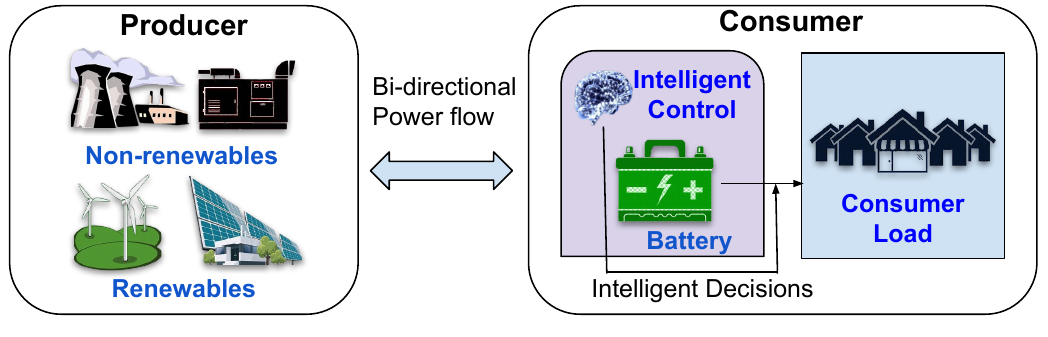}
    \caption{Overview of Energy Arbitrage Scenario.}
    \label{fig:EA}
\end{figure}

As described in Section~\ref{sec:motivation_eg}, our partner is deploying Li-ion batteries at their consumer premises for the following objectives (see Figure~\ref{fig:EA}), \textbf{(i) maximize economic profit (cost savings):} exploit price variations to increase savings by charging/discharging the battery at off-peak/peak price periods;
\textbf{(ii) maximize environmental impact (carbon footprint):} exploit generation mix (typically a combination of renewable and non-renewable sources are used to produce electricity) to reduce overall carbon footprint by charging/discharging the battery based on usage of renewable/non-renewable energy sources; \textbf{(iii) maximize both economic profit and environmental impact.} Thus, it is important to optimize the management and scheduling of these batteries, specifically, when to charge or discharge to maximize the given objective.

EA is emerging as a very crucial scenario to address the hour-to-hour variability in wind and solar, especially given the rapid increase in their adoption~\cite{vpp-survey,renewables2021iea}. EA can be framed as a maximization problem for cost and carbon savings. There exist several point solutions~\cite{battery_drl,batteries_grid,batteries_degradation1}, that have mapped this to traditional optimizations such as Markov Decision Process (MDP)~\cite{battery_drl} and Mixed Integer Linear Programming (MILP)~\cite{batteries_grid}. However, these are developed in silos, cannot be reused, and developing a solution from scratch is cumbersome and time-consuming. 
We now describe how \pname is employed to swiftly compose, analyze and derive optimal battery schedules. We describe this process by going through the three core layers. 

\subsubsection{Abstraction Layer}
\label{sec:EA_ABS}
First, the energy operator determines the entities involved in this scenario and uses the framework provided abstractions (see Section~\ref{sec:entities}). In EA there are two entity types, 

\textbf{Battery entity:} We inherit the storage class to define a Li-ion battery entity. This includes the following core configurations: (i) battery initialization parameters such as charging rates, degradation rates, state of charge (SoC), capacity, and depth of discharge, etc., (ii) $get\_state$ returns {the state of the battery} i.e current SoC of the battery, (iii) $get\_action\_space$ defines the possible actions a battery entity can take. In this scenario, we define three battery actions: \textit{charge at max rate, discharge at max rate or stay idle.} The energy operator populates these values based on their battery configuration. In {Section~\ref{sec:setup} we describe the battery configuration values used.} 

\textbf{Battery degradation:} Since batteries perform a limited number of cycles during their lifetime, we consider an accurate battery degradation model to model the battery's lifetime. We use a degradation coefficient which gets updated based on the capacity difference and the discharging cycles over a certain degradation period~\cite{battery_drl}.

\textbf{Simplified real-time market entity:} Since EA does not require any bidding decisions in the market, we modify the real-time market entity to a simplified real-time market entity that captures just the real-time market prices along with the carbon footprint information. This shows the utility of our abstractions, which allow seamless modification/extension of the definitions based on the scenario. 
\subsubsection{Environment Layer}
\label{sec:EA_ENV}
In this layer, the operator composes the actual EA scenario by defining the contract among the entities along with the decision unit to determine the optimized actions per entity.

\textbf{Decision Unit:}  In this scenario, the decision unit includes the contract between the battery and the market entity. {The decision unit here formulates state space as SoC of all involved storage entities and forecast price data of the market entity.} In EA, we do not consider any penalty with the assumption that the consumer demand will always be met. Further, the decision unit orchestrates the actions that needs to be taken for the entities. The actions determine what the battery entity should do (i.e., either charge or discharge or stay idle) and also updates battery parameters such as SoC, degradation, etc., based on the action taken. 
\subsubsection{Optimizer Layer}
\label{sec:EA_OPT}
EA scenario is formulated as a {multiple objective optimization problem}, with weights associated to maximize different objectives wrt cost savings, carbon footprint, and degradation. \textit{Thus, at every timestep the agent decides whether it should charge the battery or discharge the battery or do nothing based on the forecasts of price and carbon footprints.} The operator can now run either cost or carbon or both optimizations using any of the supported algorithms to derive the optimal battery schedules and analyze the overall savings. 
We now describe the overall methodology for EA. \\
\textbf{(i) Uncertainties involved:} In EA, uncertainties are involved in estimating future energy market prices and carbon footprint. This is very crucial information, which the optimizers will use to derive optimal schedules. As described in Section~\ref{sec:infra_components} we support multiple forecast data with different error rates to perform detailed analysis. \\
\textbf{(ii) Optimization:} The objective in EA is to either maximize cost savings or carbon footprint or both by taking into account battery degradation cost. We define $R_{t}^{price}$, $R_{t}^{carbon}$,  $R_{t}^{deg}$ as the cost, carbon footprint savings and degradation reward for a particular timeslot,
 \begin{equation}
 \footnotesize
        R_{t}^{price} = p_t*(C_t Pmax_{ch}+D_tPmax_{dis}) * \Delta t
    \end{equation}
    \begin{equation}
     \footnotesize
        R_{t}^{carbon} = c_t*(C_t Pmax_{ch}+D_tPmax_{dis}) * \Delta t
    \end{equation}
    \begin{equation}
     \footnotesize
        R_{t}^{deg} = \alpha_t*|C_t Pmax_{ch}+D_tPmax_{dis}| * \Delta t
    \end{equation}
\normalsize    
where $p_t$ and $c_t$ denotes the scaled actual price values and actual carbon footprint values, $Pmax_{ch}$ and $Pmax_{dis}$ are the maximum battery charging ($Pmax_{ch}$ < = 0) and discharging ($Pmax_{dis}$ > = 0) power values,  $C_t$ is the binary charging action taken at time t and  $D_t$ is the binary discharging action taken at time t. $\Delta t$ signifies the time for which the battery delivers the  power continuously. $\alpha_t$ is the degradation coefficient which gets updated based on the capacity difference and the discharging cycles over a certain degradation period.

Thus, the overall objective $R_{net}$ can be written as:
\begin{equation}
\label{eq:reward_ea}
\footnotesize
    R_{net} =  max \sum_{t=1}^{T}(\omega_{carbon}R_{t}^{carbon} + \omega_{price}R_{t}^{price} - \omega_{deg}R_{t}^{deg})
\end{equation}
where $\omega_{carbon}, \omega_{price}$ denotes weights of the carbon and price optimization, respectively. $\omega_{deg}$ denotes the importance given to the degradation in the battery model.

The optimizers need to abide by certain constraints to update the battery SoC. If the resulting charging/discharging action decisions do not follow these, a hard penalty is observed. 
Eq.~\ref{eq:maxsoc} validates if the formulated SoC for the next timestamp is within the defined limits and also maintains the SoC at the end of the day to an SoC level greater than $SoC_{min}$. Eq.~\ref{eq:changesoc} and~\ref{eq:updatesoc} updates the SoC level for the next timestamp. 
\begin{equation}
\footnotesize
    1-DoD <= SoC_{t+1} <= SoC_{max}; ~~~~~~~SoC_{T+1} >= SoC_{min}
\label{eq:maxsoc}
\end{equation}
\begin{equation}
\footnotesize
    SoC_{t+1} = SoC_t - \Delta SoC
\label{eq:changesoc}
\end{equation}
\begin{equation}
\footnotesize
    \Delta SoC = (\eta_{ch} C_t Pmax_{ch} + \frac{D_t}{\eta_{dis}} Pmax_{dis} ) \frac{\Delta  t}{S_t}
\label{eq:updatesoc}
\end{equation}
where $SoC_{max}$ is the maximum achievable SoC of the battery, $SoC_{min}$ is the minimum SoC that needs to be maintained at the end of the day, $DoD$ or depth of discharge represents percentage of battery capacity discharged, T denotes the end of the day and, $\eta_{ch}$ and $\eta_{dis}$ are the charging and discharging efficiencies of the battery, $S_t$ represents the current storage capacity, $SoC_t$ is the current state of charge of the battery and $SoC_{t+1}$ is the state of charge of the battery  for the immediate next timestamp. Our optimizers such as SA, MILP, and DQN-RL use the above objective function and constraints definitions to derive optimal battery schedules {(see Section~\ref{sec:results}).}  

\subsection{Scenario 2 - Optimizing Demand-side Strategies in a Micro-grid (MG)}
\label{sec:MG}
A microgrid (MG) is a local, self-sufficient energy system that allows you to generate your own electricity along with control capabilities~\cite{mopt}. Within microgrids are a group of interconnected loads (consumer demand) and distributed energy resources such as solar and wind to generate power. In addition, newer microgrids also include energy storage, typically batteries~\cite{sch}. 
When the grid goes down or electricity prices peak, MGs respond. 
In this scenario, we consider each MG is associated with a solar farm, battery storage, and consumer demand (an industrial setup). \textbf{The objective is to efficiently utilize the MG resources to satisfy consumer demand while maximizing cost savings.} In other words, the optimizing agent should derive optimal schedules to determine when to use power from the MG (energy from solar and battery) or the utility grid to match consumer demand.

MG is one of the popular new-age scenarios with different optimization objectives such as achieving the lowest prices, cleanest energy, or greatest electric reliability, etc. There are numerous challenges that need to be tackled in terms of management, planning, and
optimal use of MGs. 
We now describe how \pname is used to define and compose an MG scenario.
\subsubsection{Abstraction Layer}
\label{sec:abs_MG}
\pname supports all the entities present in an MG. Specifically, in this scenario, MG is composed of Solar, Battery, and Consumer entities, and the grid is represented with a simplified real-time market for energy pricing. 

\textbf{Solar entity:} This is inherited from the source class and includes the following configurations: \textit{(i) maximum capacity:} Total capacity of the solar farm, \textit{(ii) current capacity:} solar generation in a particular time slot, and \textit{(iii) solar profile data:} provides solar generation data either actual or forecast. \\
\textbf{Consumer entity:} This abstraction includes the $demand\_forecast$ and $demand\_actual$ attributes. \\
Battery and simplified real-time market abstractions remain the same as described in Scenario 1.
\subsubsection{Environment Layer}
\label{sec:env_mg}
\textbf{Decision Unit:} Unlike in scenario 1, where there exists just one contract between market and battery, here there are several contracts namely between, real-time market and MG, consumer and MG, MG and solar farm, and MG and battery. 

The decision Unit takes in all the contracts between the defined entities as input and determines the action space for every timestep. The action space includes battery charge/discharge.
\subsubsection{Optimizer Layer}
\label{sec:opt_mg}
The MG scenario is formulated as a maximization problem of economic profits for the consumer. Thus, at every timestep the agent decides whether it should charge the battery or discharge the battery or do nothing based on the forecasts of price, solar generation, and consumer demand. 
We now describe the overall methodology:\\
\textbf{(i) Uncertainties involved:} The uncertainties are involved in solar generation data, market price, and consumer demand. \\
\textbf{(ii) Optimization:} The optimizers take care of the uncertainty in the forecasts by using the forecast values when training and employing the actual values to get the rewards on the objective. They learn the relation between the actuals and forecasts and generate optimal decisions for the MG. The objective in this scenario is to maximize profits for the consumer. 
Mathematically, we define $R_{net}$ as the price savings for a particular time slot, $ R_{net} = max \sum_{t=0}^{T}-(Price_{t}E_{t}^{Ugrid})$, 
 Decisions within the framework are
based on data received from forecasting models/services. We
have developed a unified forecasting module, that supports numerous state-of-the-art time-series forecasting techniques and
also operators can bring their own forecaster model and integrate it. We support a wide range of techniques from simple
5
heuristics to state-of-the-art algorithms, i.e., (i) Noise forecast: gaussian noise is added to actual data , (ii) Yesterday’s
forecast: previous day’s actual data as forecast for the next
day, (iii) Mean: mean of actual data for previous N days as
forecast, (iv) LightGBM [25], a decision tree based algorithm
that builds a tree-like structure to perform regression and (v)
N-BEATS [31], a purely neural network-based model.
where $Price_{t}$ is the scaled price value between [-1,1] and $E_{t}^{Ugrid}$ is the energy supplied by the main Utility Grid. The constraints capture battery SoC updates (similar to scenario 1) and ensure instantaneous energy conservation i.e., the total energy supplied to the consumer should be equal to the total energy consumed every timeslot.
Note that, one can easily extend scenario 1 to define the MG scenario (scenario 2) by adding additional entities and constraints, owing to the general abstraction, extensible, and composable design of \pname.  


\subsection{Scenario 3 - 24/7 Demand Matching and Bidding Optimization (BO)}
\begin{figure}[t]
    \centering
\includegraphics[width=0.9\columnwidth]{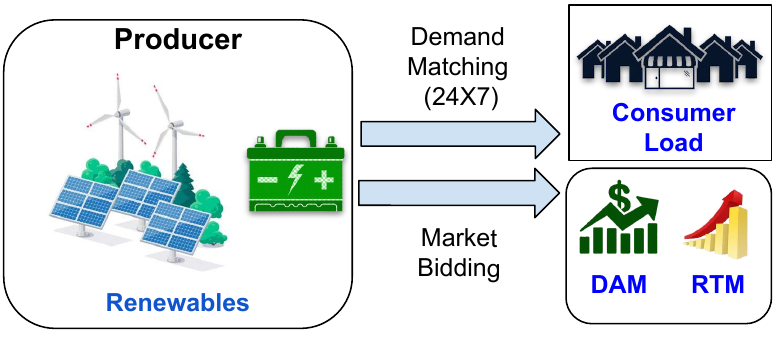}
    \caption{Demand matching and Bidding overview. }
    \label{fig:sc2-diagram}
\end{figure}

The primary challenge with renewable sources is the uncertainty in the forecast for energy generation. Also traditionally, this uncertainty has existed at a longer time frame (a few days) due to supply chain dependencies in raw materials such as coal. This shift from a longer time frame uncertainty to higher uncertainty in a shorter time horizon has had several market implications. The market structure to support these uncertainties among others includes 
real-time and day-ahead markets, which offer the sale and purchase of energy for each hour of the following day.
Further, for maintaining contracts with consumers with their own typical demand curves, there is also a need for use of storage sources - e.g., Li-ion batteries for a smaller time frame and Pumped Hydro Energy Storage (PHES) for a longer time frame. 

As shown in Figure \ref{fig:sc2-diagram}, we demonstrate a situation where weather-dependent sources - solar and wind - along with a PHES storage model are used to (i) meet the demands of a consumer contract with daily matching of supply-demand, (ii) monetization of excess energy available from the sources by bidding into energy markets. We consider two market scenarios, \textit{viz.,} day-ahead (DAM) and real-time markets (RTM). 

In the presence of the uncertainty, bidding into the multiple available markets based on the price and generation forecasts, and operating the battery storage systems has become a non-trivial operations problem for operators.
%
While many bespoke solutions~\cite{wang2020day,fazlalipour2019risk} have been proposed in the literature, our primary distinction comes from the flexibility that the framework provides in defining and optimization for the scenario using the layered execution structure discussed in Section~\ref{sec:architecture}.
In comparison, our framework automatically provides support for the bidding interval identification based on market definition and incorporates the contract penalty. The committed bids and the corresponding penalty variables are maintained as system states by the framework in the environment layer and can be accessed directly by the optimizer.
\subsubsection{Abstraction layer}
This scenario includes \textit{two entity sources} - solar and wind, which simply generate power based on a forecast. A \textit{battery entity} operates on fixed charging/discharging power. The decision has to be taken for every 15 mins timestep on whether the battery would charge/discharge at this time. The abstraction layer also contains \textit{two market entities} - a day-ahead market entity has a bid window of 2 hours from 12 noon - 2 pm every day, where 96 slots of 15 mins each for the next day have to be bid. The real-time market (RTM) has a bid window every half hour where bidding for 2 slots 1-hr ahead of the current time is made. The two markets are used to monetize the excess power generated. A safer bid is made in the DAM market and then the balance is operated with the real-time market.
\subsubsection{Environment Layer}
\textbf{Decision Unit:} In this scenario, there exist contracts between the sources, markets, and batteries. The entity formulation layer of the markets contains decision variables/actions as per the number of slots to bid for. The \textit{update} routines update the number of values that are committed, scheduled, or delivered. Based on this information, the contract penalty and the DSM penalty are estimated and used for optimization. The optimization slots for which the models must be built for all entities before composition are determined through the decision unit. For example, at 2 pm, the decisions taken, affect the 96 slots for the next day as well as the slots for 3, 3:15 pm in the same day. The environment automatically identifies these as non-overlapping decisions and builds models for the current day and the next day appropriately. The composed environment layer now exposes the revenue, penalty, and transmission penalty variables to the optimization layer.
\subsubsection{Optimization Layer}
The optimization layer maximizes the profit for the producer based on the exposed revenue, penalty, and variables from the environment layer for every set of optimization times. The uncertainties include price variations in DAM and RTM markets, and energy generation from solar and wind. 

Market bidding optimization is a prevalent new-age energy scenario and we showed that the modular design of \pname framework enables the composition of such complex scenarios with multiple sources and markets.
\section{Experimental Setup \& Implementation}
\label{sec:setup}
{We implemented \pname, including core components (entities, contracts, decision units) and core layers (abstraction, environment, optimizers) in >10K lines of python code.} The codebase along with ready-to-consume notebooks is made available\footnote{\pname \repo}. We will now describe the datasets and configurations used to evaluate the three scenarios.

\noindent\textbf{1. Energy Arbitrage Scenario}

\textit{Datasets:} We use public datasets from UK and US (California) to obtain hourly price (in euros or dollars) and carbon footprint (in gCO2eq) information. The UK data is obtained from National Grid ESO~\cite{ukdata} for the years 2018, 2019, and 2020, and US data from CAISO~\cite{caisodata} for 2018 and 2019.  

\textit{Entity configurations:} The battery configurations obtained by our industry partners are below. For UK setup, we use a 10kWh capacity, of efficiency 1.0, with a depth of discharge of 90\%, having an initial state of charge of 0.5. For CAISO, we use 50kWh capacity with an efficiency of 1, having an initial state of charge of 20\%, with a 100\% depth of discharge. 

\noindent\textbf{2. Microgrid Scenario}

\textit{Datasets:} We use the public Schneider Electric Microgrid dataset to evaluate scenario 2. The dataset includes 70 industrial sites data, with each site having a solar farm, battery, and consumer demand data every 15 minutes. The dataset also provides battery configurations per site along with electricity price information. More details can be found here~\cite{sch}.
\textit{Entity configurations:} We use the same configurations provided in the dataset to populate solar, battery, market, and consumer entity abstractions. 

\noindent\textbf{3. Market Bidding Optimization}

\textit{Datasets:} We use India Energy Exchange (IEX) dataset~\cite{iexdata} to obtain day-ahead (DAM) and real-time markets (RTM) price information. 

\textit{Entity configurations:} The source and battery configurations are similar to scenarios 1 and 2. The market configurations are similar to IEX regulations ~\cite{iex_reg}.

\section{Results}
\label{sec:results}
We now present detailed results to showcase the core features of \pname framework to support new-age scenarios.
\subsection{Support for composabilty}
\label{sec:result_abs_comp}

\begin{table}[t!]
\centering
\caption{\footnotesize{Variations and complexity of \pname for pre-defined scenarios with E= Energy Source, S= Storage, M= Market, C= consumer demand.}}
\label{tab:abs-comp}
\resizebox{0.5\textwidth}{!}{%
\begin{tabular}{|c|c|c|c|c|c|c|c|}
\hline
Scenario &
  E &
  S &
  M &
  C &
  Env &
  \begin{tabular}[c]{@{}c@{}}Max actions\\ \end{tabular} &
  \begin{tabular}[c]{@{}c@{}}State space\\ \end{tabular} \\ \hline
Scenario 1- BA             &\---               & \checkmark        & \xmark      &\---           & Fixed       & $S$ & $H$             \\ \hline
Scenario 2 - MG       & \checkmark             & \checkmark        & \xmark      & \checkmark         & Fixed       & $S + \frac{M*T}{d}$    & $H$      \\ \hline
Scenario 3- BO & \checkmark             & \checkmark        & \checkmark       & \checkmark         & Dynamic     & $S + \frac{M*T}{d} + C$  & $H*(1+M+C)$    \\ \hline
\end{tabular}
}
\end{table}
To demonstrate the power of abstractions and composability of \pname, we have developed a few real-world pre-defined scenarios using the framework as discussed in Section \ref{sec:scenario_overview}. Table \ref{tab:abs-comp} demonstrates the variations and increasing level of complexity of these scenarios. As shown in the Table, the different scenarios compose different combinations of Energy Sources- like solar, wind etc., Storage like battery vs. hydro storage, different forward bidding markets, and consumer side participation or control. Also, for the scenarios of battery arbitrage and Microgrid optimization, there is single optimization is performed  per time step. However, in Market bidding scenario, there are multiple levels of cascading optimization decisions that needs to be performed for committing to the day-ahead, intra-day and real-time markets, battery actions and allocation decisions (dynamic environment), thereby expanding the state-space and action space of decisions. This is illustrated by the state and action space sizes per Scenario. $H$ represents the historical time intervals data included in the state space, $T$ indicates the time interval over which decisions are made, and $d$ indicates the optimization time step. 

\textbf{Summary:} \pname can be used to quickly formulate and compose increasingly complex scenarios efficiently.

\subsection{Support for different optimizers}
\label{sec:result_diff_opt}
Table~\ref{tab:diff_opt} shows the savings obtained for all three scenarios using the optimizers supported by \pname Optimizer layer (discussed in Section \ref{sec:overview}) with forecast computed using the mean of the previous 10 days' data. For scenario 1, we used 1 year of test data (year 2020) across all optimizers and for DQN-RL we used 2 years of data (years 2018-19) for training. The result shows over 88K EUR in cost and 163 $kgCO_2eq$ carbon reduction can be achieved over a period of one year, which is equivalent to greenhouse gas emissions from 35000+ gasoline-powered passenger vehicles driven for one year. For scenario 2, we use 305 days of test data and 455 days of train data for DQN-RL. We show average cost savings of 4738K EUR across all 70 sites over the test period. {For scenario 3, we use 1 year of day-ahead market data (year 2020) from IEX Market and use MILP to generate cost savings for a renewable farm. The results show a net income of 581.3M INR using MILP, with a profit of approximately 20.4M INR as compared to a heuristics baseline, which  assigns volumes to markets with a greedy strategy to maximize profits after meeting certain consumer fulfillments without looking for the longer contract reward}. 
In general, both MILP and DQN-RL have similar performances. In the next section, we show how the optimizer's performance changes as the forecast errors increase. Note, for the remaining sections, we will focus on Scenario 1 - energy arbitrage results and similar results can be derived using our framework for scenarios 2 and 3.  

\textbf{Summary:} With native support for different optimizers \pname allows energy operators to analyze the decisions without worrying about the implementation details. 

\begin{table}[t!]
\centering
\caption{Savings observed for scenarios 1-3 using different optimizers supported by \pname.}
\label{tab:diff_opt}
\resizebox{0.45\textwidth}{!}{%
\begin{tabular}{|c|cc|c|c|}
\hline
\multirow{3}{*}{Optimizers} & \multicolumn{2}{c|}{Scenario 1 - EA} & Scenario 2 - MG & Scenario 3- BO \\ \cline{2-5} 
 &
  \multicolumn{1}{c|}{\multirow{2}{*}{\begin{tabular}[c]{@{}c@{}}Cost savings\\ (K EUR)\end{tabular}}} &
  \multirow{2}{*}{\begin{tabular}[c]{@{}c@{}}Carbon savings\\ ( $kgCO_2eq$)\end{tabular}} &
  \multirow{2}{*}{\begin{tabular}[c]{@{}c@{}}Cost savings\\ ( K EUR)\end{tabular}} &
  \multirow{2}{*}{\begin{tabular}[c]{@{}c@{}}Profits\\ (M INR)\end{tabular}} \\
                            & \multicolumn{1}{c|}{}       &        &                 &                \\ \hline
SA     & \multicolumn{1}{c|}{54.5}  & 102.5 & 4732.5 & NA \\ 
MILP   & \multicolumn{1}{c|}{\textbf{88.1}}  & \textbf{163.3} & 4738 .1& {581.3} \\ 
DQN-RL & \multicolumn{1}{c|}{85.1}  & 140.2 & \textbf{4738.2} & NA \\ \hline
\end{tabular}%
}
\end{table}

\subsection{Support for uncertainty in forecasts}
\label{sec:result_forecasts}
We explore the impact of uncertainty in the forecasts while optimizing for cost and carbon savings. Using MILP and DQN-RL as our optimizers on six forecasting methods, discussed in \ref{sec:infra_components}. From Table \ref{tab:milp_dqn_for_forecasts_ba}, we observe that the forecasting errors for price forecasting are relatively low (MAE<11) for all forecasting methods compared with the forecasting errors for carbon forecasting (with MAE >40). This is due to the fact that the time-series of carbon emissions lack seasonality and is often harder to predict.
DQN-RL outperforms MILP in cost savings, owing to the training routine to learn and adapt to forecast errors. However, it performs similarly to MILP in carbon savings, mainly due to the fact that accounting for high uncertainty is a much more difficult task for DQN-RL and requires additional training data.

\textbf{Summary:} Neural network-based formulations outperform traditional algorithms in presence of forecast errors and sufficient training data to learn from. \pname provides easy support to evaluate outcomes of different optimizers on various available forecasting functions/algorithms.

\begin{table}[t!]
\caption{Cost and carbon savings on different forecasting methods with varying mean average error (MAE).}
\resizebox{0.45\textwidth}{!}{%
\begin{tabular}{|c|cc|cc|cc|}
\hline
Forecast Method &
  \multicolumn{2}{c|}{Forecasting MAE} &
  \multicolumn{2}{c|}{\begin{tabular}[c]{@{}c@{}}Cost Savings\\ (K EUR)\end{tabular}} &
  \multicolumn{2}{c|}{\begin{tabular}[c]{@{}c@{}}Carbon Savings\\ ($kgCO_2eq$)\end{tabular}} \\ \cline{2-7} 
 &
  \multicolumn{1}{c|}{Price} &
  Carbon &
  \multicolumn{1}{c|}{MILP} &
  DQN &
  \multicolumn{1}{c|}{MILP} &
  DQN \\ \hline
Accurate &
  \multicolumn{1}{c|}{0} &
  0 &
  \multicolumn{1}{c|}{116.1} &
  \textbf{117.73} &
  \multicolumn{1}{c|}{\textbf{227.27}} &
  211.59 \\ \hline
Noise &
  \multicolumn{1}{c|}{19.82} &
  19.08 &
  \multicolumn{1}{c|}{44.64} &
  \textbf{57.86} &
  \multicolumn{1}{c|}{123.93} &
  \textbf{163.17} \\ \hline
Yesterday &
  \multicolumn{1}{c|}{9.70} &
  42.19 &
  \multicolumn{1}{c|}{77.74} &
  \textbf{83.08} &
  \multicolumn{1}{c|}{\textbf{138.384}} &
  129.37 \\ \hline
Mean &
  \multicolumn{1}{c|}{9.94} &
  48.13 &
  \multicolumn{1}{c|}{\textbf{88.2}} &
  83.24 &
  \multicolumn{1}{c|}{\textbf{163.34}} &
  135.06 \\ \hline
LightGBM &
  \multicolumn{1}{c|}{11.88} &
  38.33 &
  \multicolumn{1}{c|}{55.89} &
  \textbf{62.14} &
  \multicolumn{1}{c|}{124.4} &
  \textbf{128.73} \\ \hline
N-BEATS &
  \multicolumn{1}{c|}{10.13} &
  38.49 &
  \multicolumn{1}{c|}{73.3} &
  \textbf{76.71} &
  \multicolumn{1}{c|}{\textbf{145.82}} &
  130.94 \\ \hline
\end{tabular}%
}
\label{tab:milp_dqn_for_forecasts_ba}
\end{table}

\subsection{Support for Extensibility }
\label{sec:result_compose}
Table~\ref{tab:joint_opt} demonstrates the extensibility support of \pname through ease of composition from the single objective (either cost or carbon optimization) to multi-objective (having joint cost and carbon optimization) results for scenario 1 using MILP and DQN-RL. Given the composable \& extensible formulation as shown in earlier section, it is trivial for energy operators to extend the single objective functions to multi-objective through configuration. To solve the objectives, we leverage the market price data and carbon emissions data available from the UK grid. The DQN-RL agent is then trained on 2 years (2018-19) UK data (training data) and results are inferred on 1 yr (2020) data of the same UK grid (test data). We utilize the same 1 year's test data to produce results for MILP. Here, we show results for both accurate forecasts as well as forecasts by taking the mean of previous 10 days' actual data. We observe that when the forecasts are accurate, joint optimization results in cost and carbon savings that are close to the optimization results obtained by solving individual objectives. In both carbon as well as cost savings, the joint-optimization savings are lower by approximately 30\% as compared to the respective individual objectives. Similarly with the forecast data, the numbers significantly reduce by approximately 43 - 47\% while using joint optimization as opposed to the single objectives.
In these experiments of joint optimization, we are using a weight factor ($\alpha = \frac{\omega_{carbon}}{\omega{price}}$) of 1.155 and 1.28 for MILP and RL respectively. While any number can be used as a configuration parameter for $\alpha$, to determine this optimal value we perform a detailed what-if analysis, described in the next section.

\textbf{Summary:} The savings reduce with the increase in the uncertainty of forecasts. \pname enables operators to extend existing scenarios through configurations.

\begin{table}[]
\centering
\caption{Savings comparison using single and multi-objectives for Energy Arbitrage scenario.}
\label{tab:joint_opt}
\resizebox{0.5\textwidth}{!}{%
\begin{tabular}{|c|cc|cc|}
\hline
\multirow{2}{*}{Optimizers} & \multicolumn{2}{c|}{With Accurate forecasts} & \multicolumn{2}{c|}{with Mean forecasts} \\ \cline{2-5} 
                      & \multicolumn{1}{c|}{MILP}    & DQN-RL  & \multicolumn{1}{c|}{MILP}    & DQN-RL \\ \hline
Cost savings (K EUR) & \multicolumn{1}{c|}{116.1}  & \textbf{117.7}  & \multicolumn{1}{c|}{85.1}   & \textbf{88.1 } \\ \hline
Carbon savings ($kgCO_2eq$)  & \multicolumn{1}{c|}{\textbf{227.2}}    & 211.5    & \multicolumn{1}{c|}{\textbf{163.3}}  & 140.2  \\ \hline
Joint OPT - Cost  (K EUR)    & \multicolumn{1}{c|}{74.8}   & 81.5   & \multicolumn{1}{c|}{46.2}   & 24.9  \\
Joint OPT - Carbon ($kgCO_2eq$)   & \multicolumn{1}{c|}{164.7} & 137.7 & \multicolumn{1}{c|}{135.8} & 92.6 \\ \hline
\end{tabular}%
}
\end{table}

\subsection{Support for what-if analysis}
\label{sec:result_whatif}

\begin{figure}
	\centering
	\begin{minipage}[t]{0.23\textwidth}
		\centering
\includegraphics[width=0.99\columnwidth,height=1.1in]{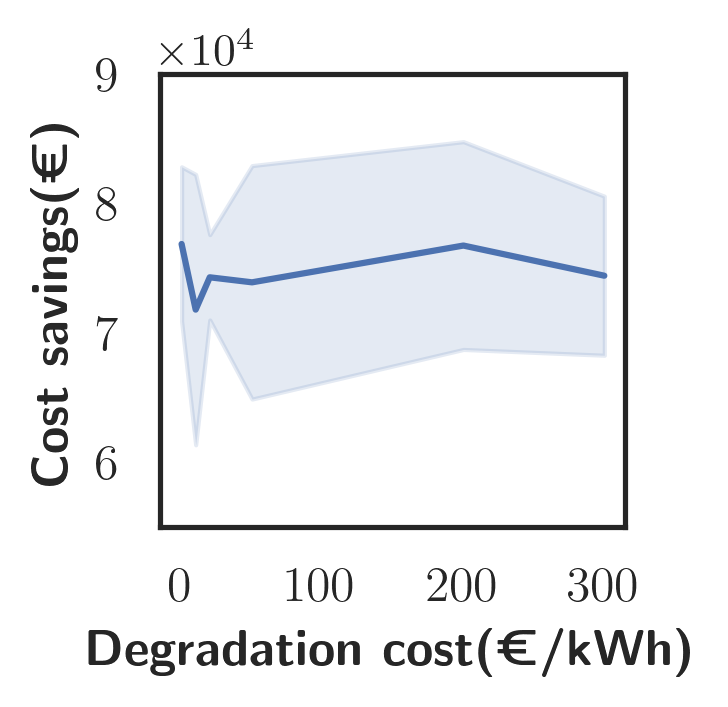}
    \caption{Savings vs Degrad.}
    \label{fig:deg_savings}
	\end{minipage}%
	\hspace{1ex}
\begin{minipage}[t]{0.23\textwidth}
		\centering
\hfill\includegraphics[width=0.99\columnwidth,height=1.1in]{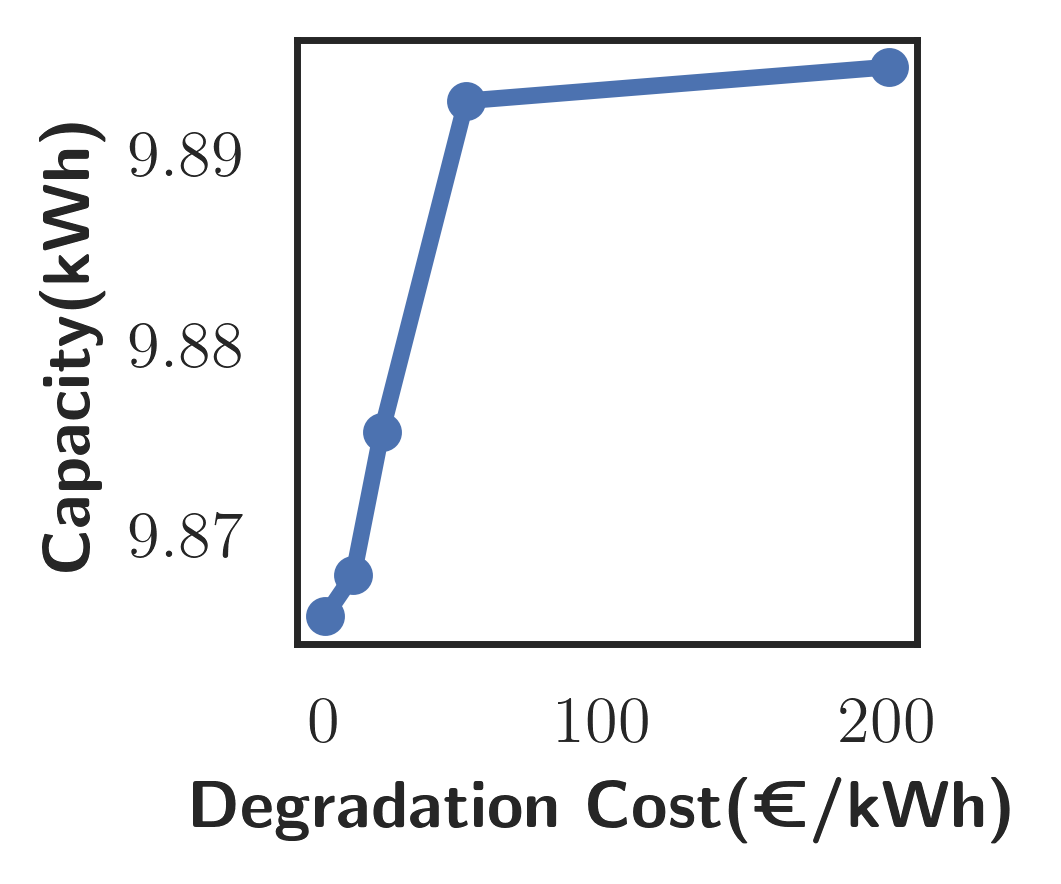}
    \caption{Capacity degrad.}
    \label{fig:deg_cap_vs_degcost}
	\end{minipage}
\end{figure}

\textbf{Pareto intersection results.}
\begin{figure}
	\centering
	\begin{minipage}[t]{0.23\textwidth}
		\centering
\includegraphics[width=0.9\columnwidth,height=1.1in]{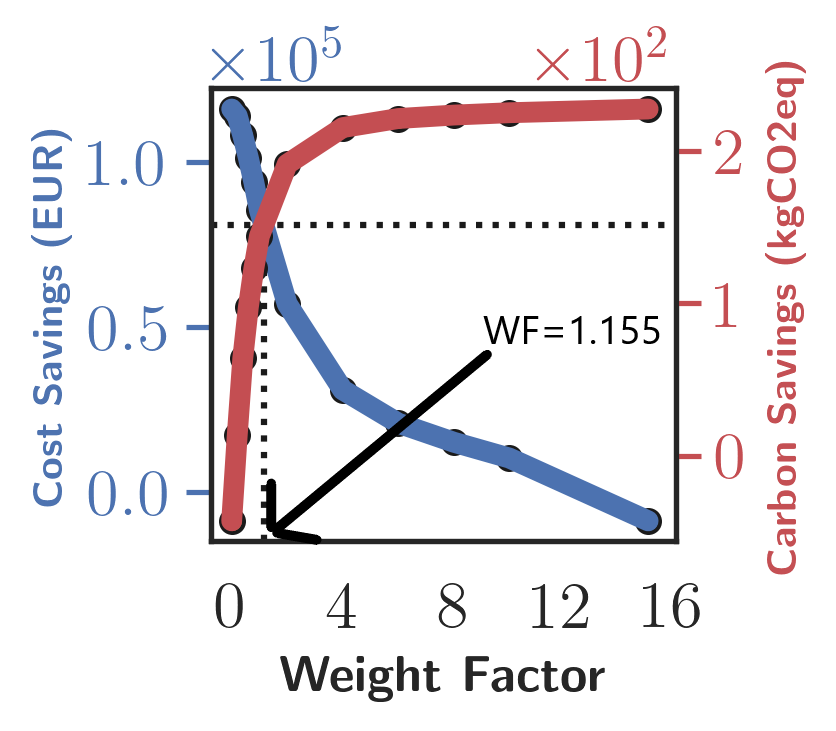}
\subcaption{\label{MILP}Pareto - MILP}
	\end{minipage}%
	\hspace{1ex}
\begin{minipage}[t]{0.23\textwidth}
		\centering	\includegraphics[width=0.9\columnwidth,height=1.1in]{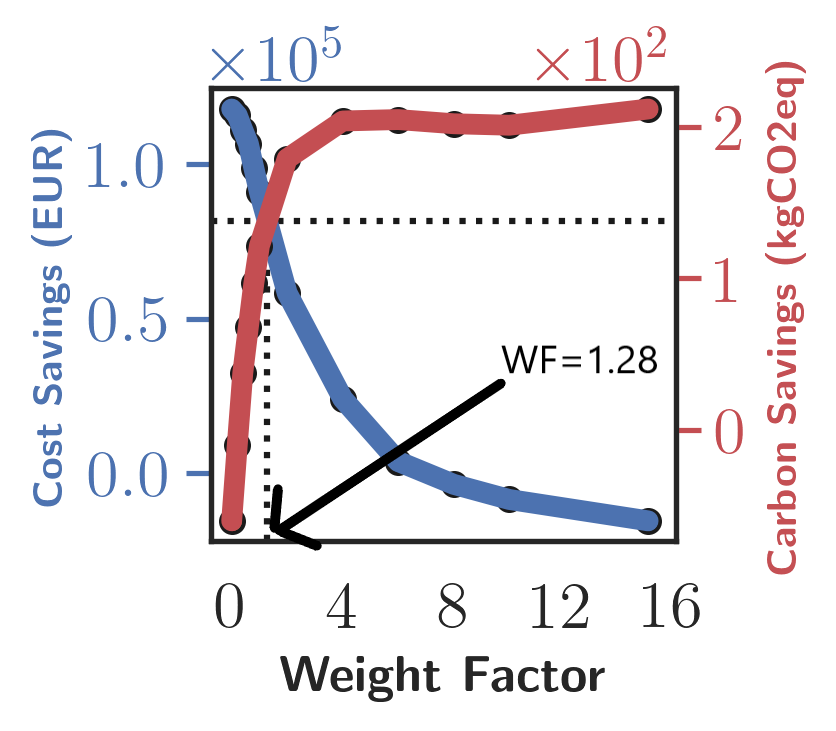}
\subcaption{\label{DQN-RL} Pareto - DQN-RL}
	\end{minipage}
		\caption[MILP DQN-RL]{Optimal points for multi-objective problem setting \label{fig:pareto}}
\end{figure}
As seen in the previous sub-section, we use an optimum value of weight factor for joint optimization of carbon and price savings. To find this optimal point, \pname supports very easy configuration-based alternatives to search through different $\alpha$ values as a hyper-parameter in one go. The user is supposed to enter the configuration values or range required to be searched for, and the system runs and provides the final optimal value. Figure~\ref{fig:pareto} shows the optimal point being determined to solve the multi-objective joint optimization of cost and carbon savings using MILP and RL respectively. The pareto-intersection of the price and carbon savings' curves varied across different $\alpha$ values to generate the optimal point. \\
\textbf{Adding degradation and varying the associated cost.}
The carbon and price savings results for battery arbitrage scenario discussed in earlier sub-sections did not include the battery degradation over multiple charging/discharging cycles. We now use degradation as a part of what-if analysis. In Figure~\ref{fig:deg_savings}, we vary the degradation cost and see how the savings change. We performed the experiment across 3 different seeds using DQN-RL and have the confidence interval plotted in the figure. The plot shows that cost savings almost remains constant when degradation cost increases. This is due to the fact that the DQN-RL optimizer learns to perform actions that doesn't quickly degrade the battery capacity since we include the degradation coefficient in the reward function (see Eq.~\ref{eq:reward_ea}). As the battery degradation cost increases, the degradation coefficient decreases. The degradation coefficient being proportional to the change in capacity, nullifies the effect of degradation. Figure~\ref{fig:deg_cap_vs_degcost} shows how the capacity degradation varies for a 10KWh capacity battery over a year with the increasing degradation cost. As we can see, there is less capacity degradation with increasing degradation cost. Further, the number of frequent charging and discharging cycles  drops by 13\% when degradation cost is considered as compared to no degradation (477 cycles vs 551 cycles). \\
\textbf{Summary:} \pname supports various types of what-if analysis for energy operators through easy configurations, in order to evaluate and determine the optimal decisions.

\subsection{Support for MLOps workflows}
\label{sec:result_mlops}

We now present a few key results showcasing the MLOps workflows supported by \pname. MLOps workflow allows energy operators to save different models and their trajectories to do what-if analysis or re-train on additional data, etc. To this end, we employ our public Azure Machine Learning (AML) service to build, deploy and perform various MLOps. 

\textbf{Expert RL agents to support lack of data:}
Often energy operators have multiple agents trained on several scenarios that are experts in their respective tasks. These expert agents with learned trajectories can be easily utilized for a new scenario. Let’s say an energy operator requires installation of their energy resources in a new region/geography where the available data is sparse. In such cases, the expert agents trained on a different dataset can be re-used in neural network-based solutions for decision making and planning.

\begin{figure}
	\centering
	\begin{minipage}[t]{0.23\textwidth}
		\centering
\includegraphics[width=1.1\columnwidth,height=1in]{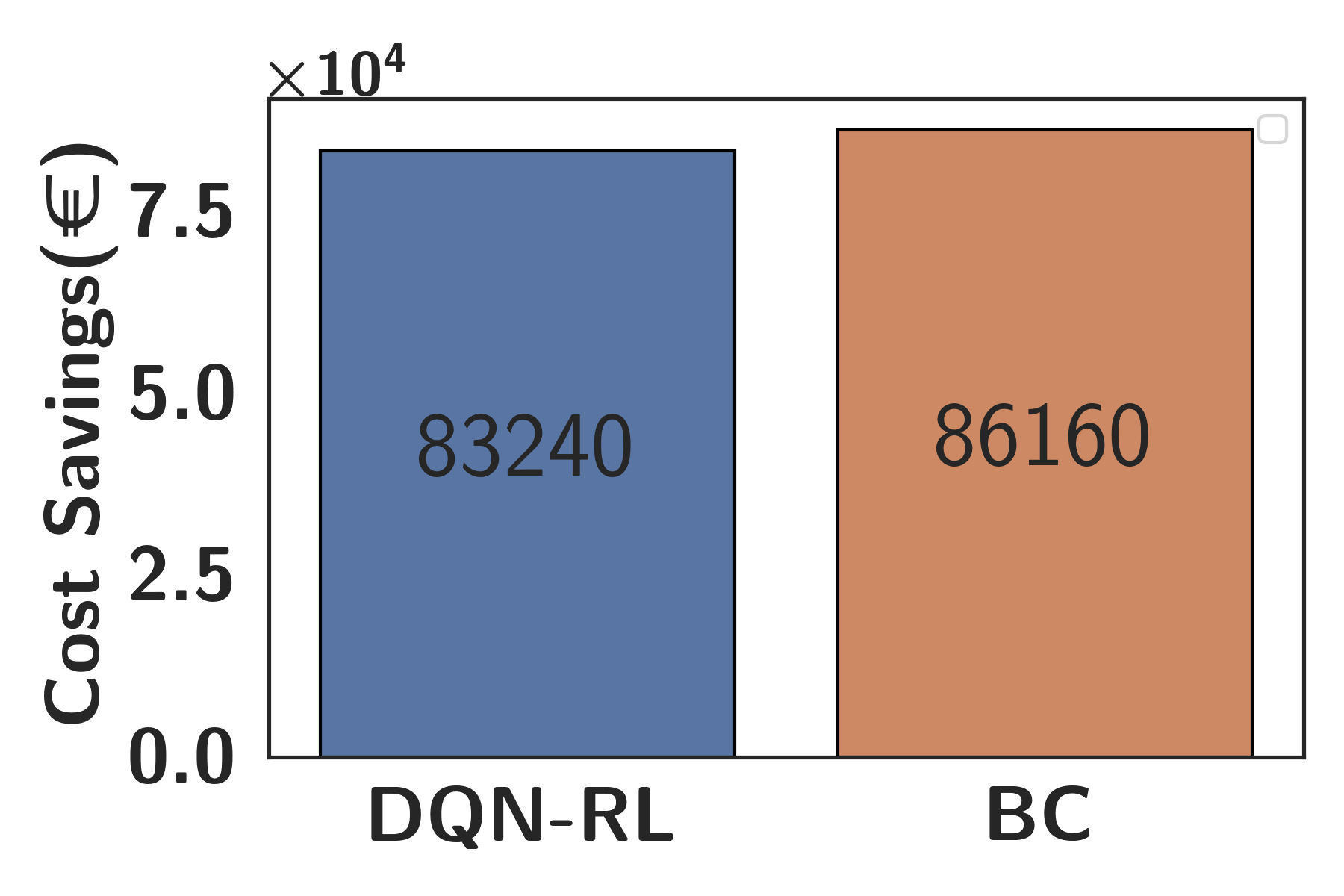}
\subcaption{Behaviour Cloning}
\label{fig:bc}
	\end{minipage}%
	\hspace{1ex}
\begin{minipage}[t]{0.23\textwidth}
		\centering
\includegraphics[width=1.1\columnwidth,height=1in]{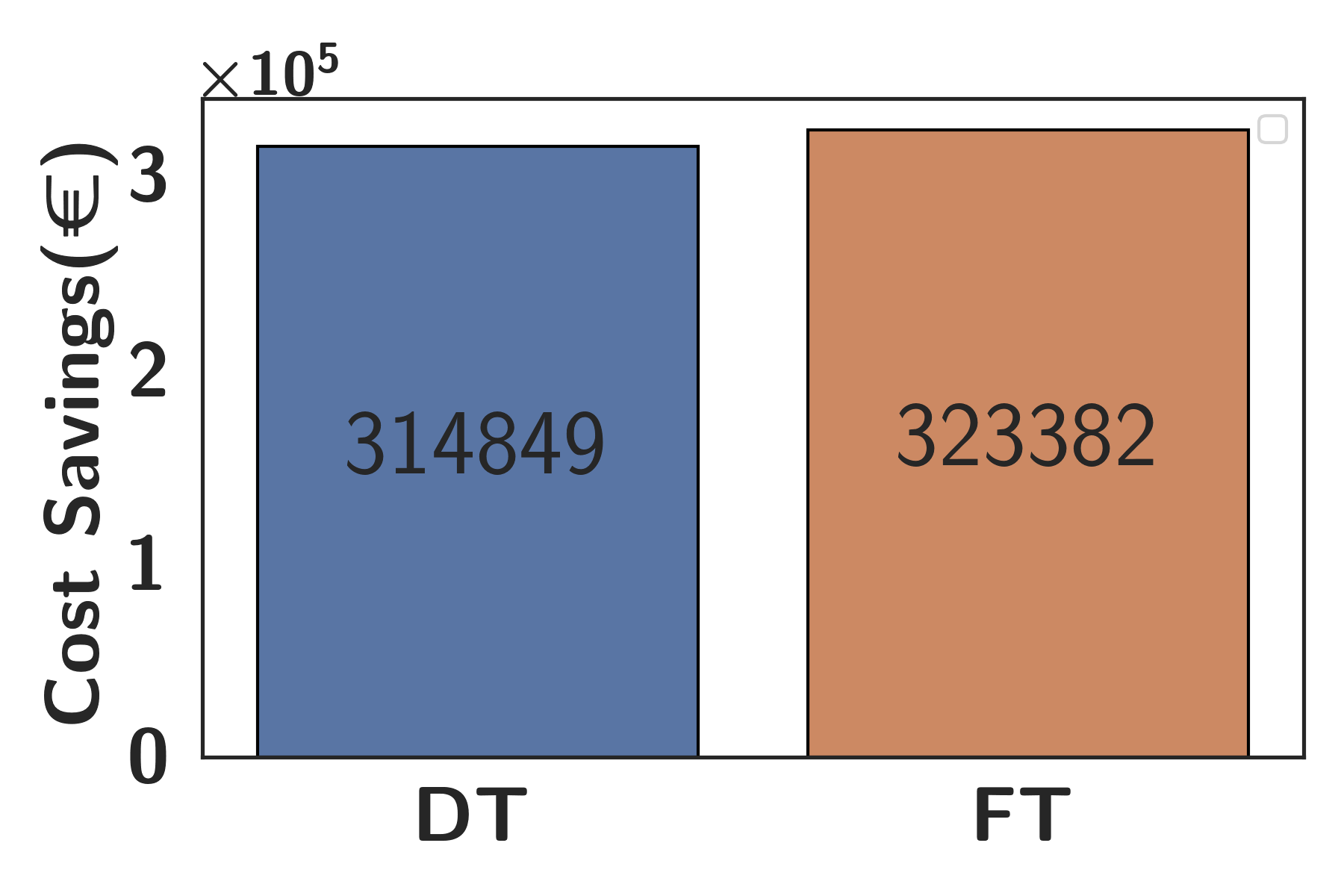}
\subcaption{ Pre-training vs Direct}
\label{fig:far_transfer}
	\end{minipage}
		\caption[MLOps Workflows]{MLOps workflows.}
  \label{fig:mlops_workflows}
\end{figure}

Specifically, the expert agent, instead of providing the child agent with the reward value functions, sets a list of demonstrations by sampling from expert transitions. The new agent tries to learn the optimal policies by imitating the expert’s decisions. 
To support this, \pname employs imitation learning library~\cite{imi} and behavior cloning (BC) models. We use 2 years of price data (2018 and 2019 from UK) to train both a DQN-RL and our expert agent. Then, we use 1-year data (2020 from UK) to show the savings obtained using DQN-RL (naive without any additional learning) and BC model (which uses information from trained expert trajectories). In Figure~\ref{fig:bc}, we can see that the BC model outperforms traditional DQN-RL, owing to the expert trajectories used to learn the observation-action pairs. 

\textbf{Use pre-trained models for new configurations and dataset:}
Energy Operators can also move from learning in one environment to another. Here, environment changes refer to testing the existing agents on a different grid with very different abstraction configurations. The intuition here is to bring out generalization across environments, even if the agents use different state representations or action spaces. 
Specifically, we use a pre-trained RL agent trained on UK data and re-train it for CAISO (US) data, and compare its performance against an RL agent that is trained only on CAISO data. Note that, the dataset size and battery configurations in these two datasets are different (see Section~\ref{sec:setup}). Figure~\ref{fig:far_transfer} shows that with pre-trained models we can improve savings by 3\% on CAISO data as opposed to direct training. The pre-trained models help energy operators especially when there is sparse data. 
Appendix~\ref{appendix:mlops} discusses other workflows supported.

\textbf{Summary:} \pname supports several MLOps workflows, which allows operators to use and analyze different datasets and configurations for their planning and operations tasks.  

\subsection{\pname latency benchmarking}
\label{sec:result_benchmark}
Within \pname, the optimizers are either exact solvers (MILP), search-based solvers (SA) or learnable optimizers (DQN-RL). We now present online latency values, which measure the speed of execution runtime on the hardware. The benchmarks are performed on a VM running AMD EPYC 7742 64-Core Processor with 996GB RAM and 4 A6000 RTX GPUs. The online latency for DQN-RL on CPU and GPU are 600$\mu s$ and 4.4$\mu s$, respectively. Further, the latency for MILP and SA are 0.42s and 195s, respectively. This benchmark provides runtime information required when choosing different optimizers for offline and real-time scenarios.



\section{Limitations and Future work}
\label{sec:discussions}
With \pname, we have taken the first step towards building a general, extensible decision management framework for energy operators. We recognize that there are a few limitations/challenges in our current framework and addressing these would be an interesting future direction. 

\textbf{Seamless support for new state-of-the-art optimizers.} \pname supports most of the traditional and neural network-based optimizers in the same environment. There may be a few new optimizers, which can be applied for energy scenarios as described in~\cite{pineda2022theseus, bian2022demand}. That said, we believe one can extend the current environments to support more optimizers with fewer lines of code using existing framework utilities. 

\textbf{Recommendation engine to determine the best optimizer.} While \pname supports running detailed what-if analysis across different optimizers, for application users it would be easier if the framework could recommend the best optimizer for a given scenario, objective, and constraints. It is non-trivial to build such a general recommendation engine as it requires significant data and decisions knowledge across scenarios and is an interesting future direction. 

\textbf{Safe-RL approaches for new-age energy scenarios.} All new-age entities such as batteries, renewable sources, etc., support control features either to charge/discharge battery, curtail generation, etc~\cite{chen2021improved}. Further, the decisions obtained in \pname are eventually sent to the physical entity hardware via their SCADA system. Any safety violation can lead to reliability issues in the power grid. For example, neglecting the crucial energy balance constraint could result in either under- or over-production leading to exceeding maximum power capacity~\cite{safe_rl}. Similarly violating minimum SoC constraints for batteries would degrade them significantly quicker than expected. Thus it is important for newer formulations to incorporate such safety constraints explicitly in the objective. 
\section{Related Work}
\label{sec:related_work}
Historically, bespoke solutions have been devised to tackle specific problems in optimizing energy systems. These methods are often times inextensible and hard to adapt to newer systems with different requirements. Previously, frameworks like \texttt{ElecSim}\cite{kell2019elecsim} and \texttt{pymgrid}\cite{henri2020pymgrid} have tried to tackle individual scenarios like market bidding optimization and microgrid optimization respectively. These frameworks provide individual instances of implementing some variants of the scenario but individually suffer some drawbacks. Firstly, it is not trivial to extend \texttt{ElecSim} to scenarios that don't involve the market, because of the rigid structure it adopts when defining the market bidding scenario. On the other hand, while \texttt{pymgrid} supports variations of a microgrid, but it is not straightforward to extend it to other types of entities and forecasts, especially entities like Market that involve complex scheduling and decision requirements. Our proposed framework addresses each of these drawbacks by providing modular, extensible entities that are customizable based on requirements. Here, entities are designed around the principle that the core components are disentangled i.e. various domain experts (domain-experts, developers, etc.) can individually customize required modules without modifying unrelated components, enabling code reuse and better functionality. 
For example, we show it is easy to add a new storage device with a completely different underlying mechanism based on a different physics model.
Furthermore, we also support composing a multitude of scenarios and optimizers to enable planning and decision-making. 

\section{Conclusion}
Renewable energy sources are non-deterministic and pose additional complexity in periodic decision-making. While there are several optimization scenarios and approaches studied in the literature, those are designed and deployed as bespoke solutions and lack generality and extensibility. The whole ecosystem (including new energy resources, markets, incentives, and business models) is evolving fast.  Hence, there is an ongoing demand for an easy-to-use system that would allow energy operators to compose, build, validate and deploy their real-world scenarios rapidly. In this work, we presented \pname, a generic, modular, extensible and scalable framework for new-age energy systems that enables data-driven decision-making and provides ease of use for development and production scale deployments. Through the evaluation on three real-world scenarios, we demonstrate the power and effectiveness of \pname. 

\bibliographystyle{plain}
\bibliography{references}

\appendix

\section*{Appendix}
\section{Entity Abstraction}
\label{appendix:abstract}

\begin{figure}[ht!]
		\centering
		\includegraphics[width=0.95\columnwidth]{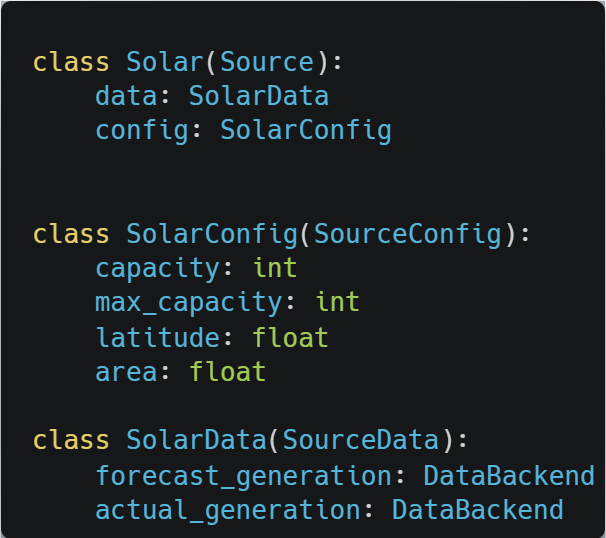}
		\caption{Solar entity abstraction}
		\label{fig:abs_eg1}
 \end{figure}

\begin{figure}[ht!]
		\centering
		\includegraphics[width=0.95\columnwidth]{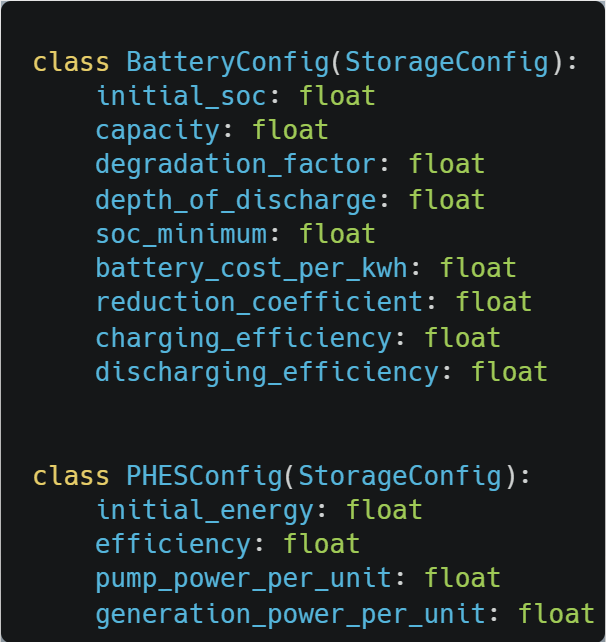}
		\caption{Battery entity abstraction}
		\label{fig:abs_eg2}
	\vspace{-15pt}
\end{figure}

Here, we briefly cover how to implement different energy sources. In Figure~\ref{fig:abs_eg1}, the \texttt{Solar} class overrides the data and config attributes to support additional configuration parameters to implement the underlying phsyics-based formulations and contains a data attribute to access various solar specific power generation attributes.  Figure~\ref{fig:abs_eg2} further illustrates how configuration parameters vary across storage devices(Battery and PHES) and the requirement of an easy-to-configure, disentangled module. 


\section{Market Definition example}
\label{appendix:market_def}
\begin{figure}
    \centering
    \includegraphics[width=\linewidth]{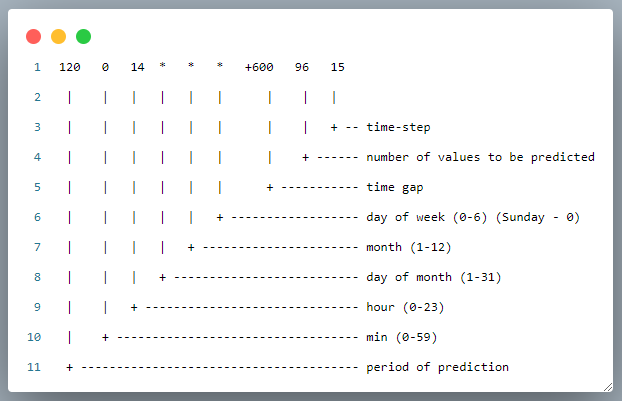}
    \caption{A cron-style schedule definition is used for defining markets that support double-blind bids. The demonstrated values are for a market with a bid window of 120 mins ending at 1400 hours where the bid must be made for 96 slots of 15 mins each starting 400 mins after 1400 hours, i.e., 0000hrs+1D.}
    \label{fig:schedule_definition}
\end{figure}

To model all these types of markets, we simplify the behavior down to a schedule definition (Figure \ref{fig:schedule_definition}). Corresponding to each market, there are periods of time where the bidding is open. The end of this window is the deadline before which certain values have to be committed to a market. Using the cron-style definition of schedule in Figure \ref{fig:schedule_definition}, we define the number of variables that has to be predicted before the deadline. For example, a day-ahead market has a bidding window from 12 noon - 2 pm every day where 96 values (bid volume, bid price) corresponding to 15-minute slots of the next day have to be defined. The schedule for this example is shown in Figure \ref{fig:schedule_definition}. The market models available in our framework currently assume that the bidders are price-takers, however, the framework supports extension of markets to a bid-sensitive price formulation. Continuous auction markets are currently not modeled and are a part of future work in the framework.\\

\section{\pname Environment Layer details}
\label{appendix:env_layer}
We now present the details of Entity and composed formulation layer by taking MILP as an example.

 \paragraph{Entity Formulation Layer: MILP Example} Let us consider an example of the layers for the MILP formulation. In this case, each of the entities expose \textit{volume\_out\_var} functions. In the case of entities like solar/wind, this is simply the generation forecast numbers. In case of a battery entity, let $C_t$ and $D_t$ represent the binary variables whether the battery is charged/discharged at the time $t$, then the entity formulation layer consists of the following:
    \begin{enumerate}
        \item \textit{model\_build}: The model build routine initializes the binary variables, imposes constraints, 
        \begin{equation}
            0 \leq C_t \leq 1, 0 \leq D_t \leq 1
        \end{equation}
        \begin{equation}
            0 \leq C_t + D_t \leq 1    
        \end{equation}
         Since the model is built for $t \in [1, T]$, this function also specifies the operating rules, i.e., 
        \begin{equation}
            E_t = E_i + \sum_{t=1}^{t}(\eta_C C_t P_C - \frac{1}{\eta_D} D_t P_D)
        \end{equation}
        where $\eta_C$ and $\eta_D$ are charging and discharging efficiencies, $P_C$ and $P_D$ are the charging/discharging powers along with their constraints, e.g., 
        \begin{equation}
            SOC_{min} * cap \leq E_t \leq SOC_{max} * cap
        \end{equation}
        where $SOC_{min}$ and $SOC_{max}$ are based on the operational usage recommendations from the battery manufacturer.
        
        \item \textit{volume\_out\_var}: 
        \begin{equation}
            \text{volume\_out\_var} = (P_D D_t - P_C C_t)
        \end{equation}
        where $P_D$ and $P_C$ are the discharging/charging power respectively.
        
        \item \textit{update}: An update routine interfaces with the underlying abstraction to update the underlying abstraction with solutions of the current optimization. For example, in this case, the battery state of charge is updated int the battery abstraction.
        \begin{equation}
            batt.soc_t = soc_{t-1} + (\eta_C P_C C_t - \frac{1}{\eta_D} D_t P_D)
        \end{equation}
    \end{enumerate}
    
    \paragraph{Composed Formulation Layer: MILP Example} With all the entities exposing the same interface, it is possible to write a generic composed formulation layer. Let $\mathbf{D}$ be the decision unit under consideration. Let $\mathbf{E}$ be the set of entities and $\mathbf{C}$ be the set of contracts in the decision unit. Let $c.C_r$ be the contractors for contract $c$, which is a list of sources for a particular contract and $c.C_e$ be the contractee for the contract $c$, i.e., market or consumer. Then, 
    \begin{equation}
        \text{revenue}_t = \sum_{c \in \mathbf{C}}(\sum_{s \in c.C_r} s.vout_t)*c.C_e.price_t
    \end{equation}
    \begin{equation}
        \text{penalty} = \sum_{c \in \mathbf{C}}[c.penalty\_fn((\sum_{s \in c.C_r} s.vout_t), c.C_e.cv_t)]
    \end{equation}
    where $vout$ and $cv$ are volume\_out (as obtained from the volume\_out\_var function in the entity formulation layer) and committed volumes respectively. The penalty\_fn is a user specific function per contract based on which the differences in committment vs supply are penalised. In the framework, we provide provisions for accuring this for different time intervals, e.g., daily or for a fixed subset of hours per day.

\section{\pname supported optimizers}
\label{appendix:opt}
\pname supports the following optimizers:\\
(i) \textit{\textbf{Simulated Annealing (SA)}} is a traditional probabilistic optimization technique, which approximates the global optimum for a given function. \\
(ii) \textit{\textbf{Mixed Integer Linear Programming (MILP)}} is also one of the traditional optimization techniques, which solves a given objective based on certain constraints over a fixed given Horizon. \\
(iii) \textit{\textbf{Model Predictive Control (MPC)}} is also a traditional technique for optimization which, unlike MILP, solves a sequence of multistage deterministic optimization problems over a given fixed horizon H. \\
(iv) \textit{\textbf{Deep Q Network based Reinforcement Learning (DQN-RL)}} is a neural network-based algorithm. The environment, agent, state, and reward form the building blocks for any RL problem. The agent interacts with the environment by taking actions in the current state to proceed to the next state. Agents get rewarded based on the actions taken in the environment. They learn how to navigate through the environment based on the experience gained to reach the end goal by maximizing the cumulative reward for every episode. DQN, being model-free, solves the RL task directly using the samples provided by the environment, without constructing explicit estimates of them.

\section{\pname supported MLOps workflows}
\label{appendix:mlops}
\begin{figure}
	\centering
	\begin{minipage}[t]{0.23\textwidth}
		\centering
\includegraphics[width=1.1\columnwidth,height=1.2in]{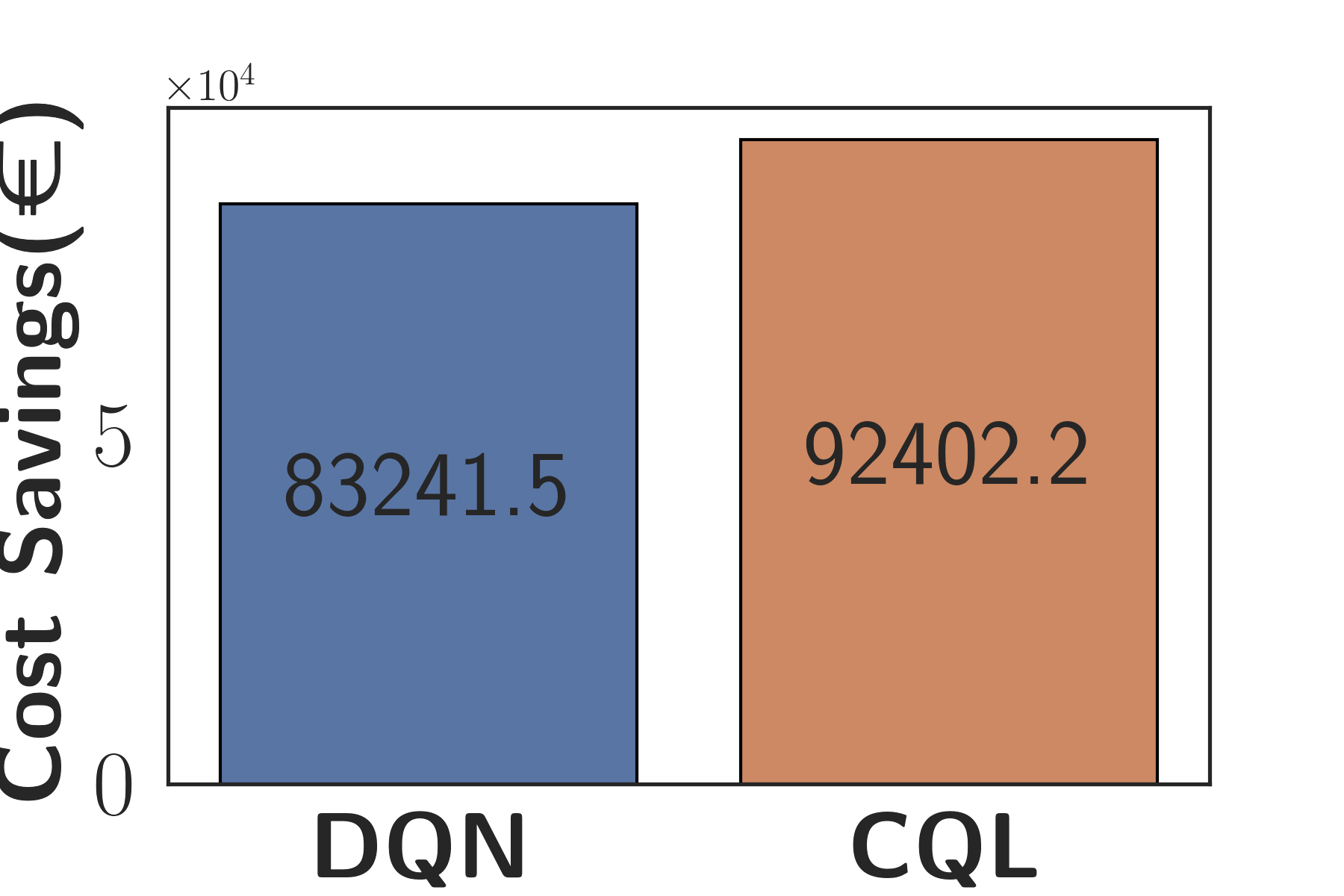}
\subcaption{Offline RL}
\label{fig:offline-rl}
	\end{minipage}%
	\hspace{1ex}
\begin{minipage}[t]{0.23\textwidth}
		\centering
\includegraphics[width=1.1\columnwidth,height=1.2in]{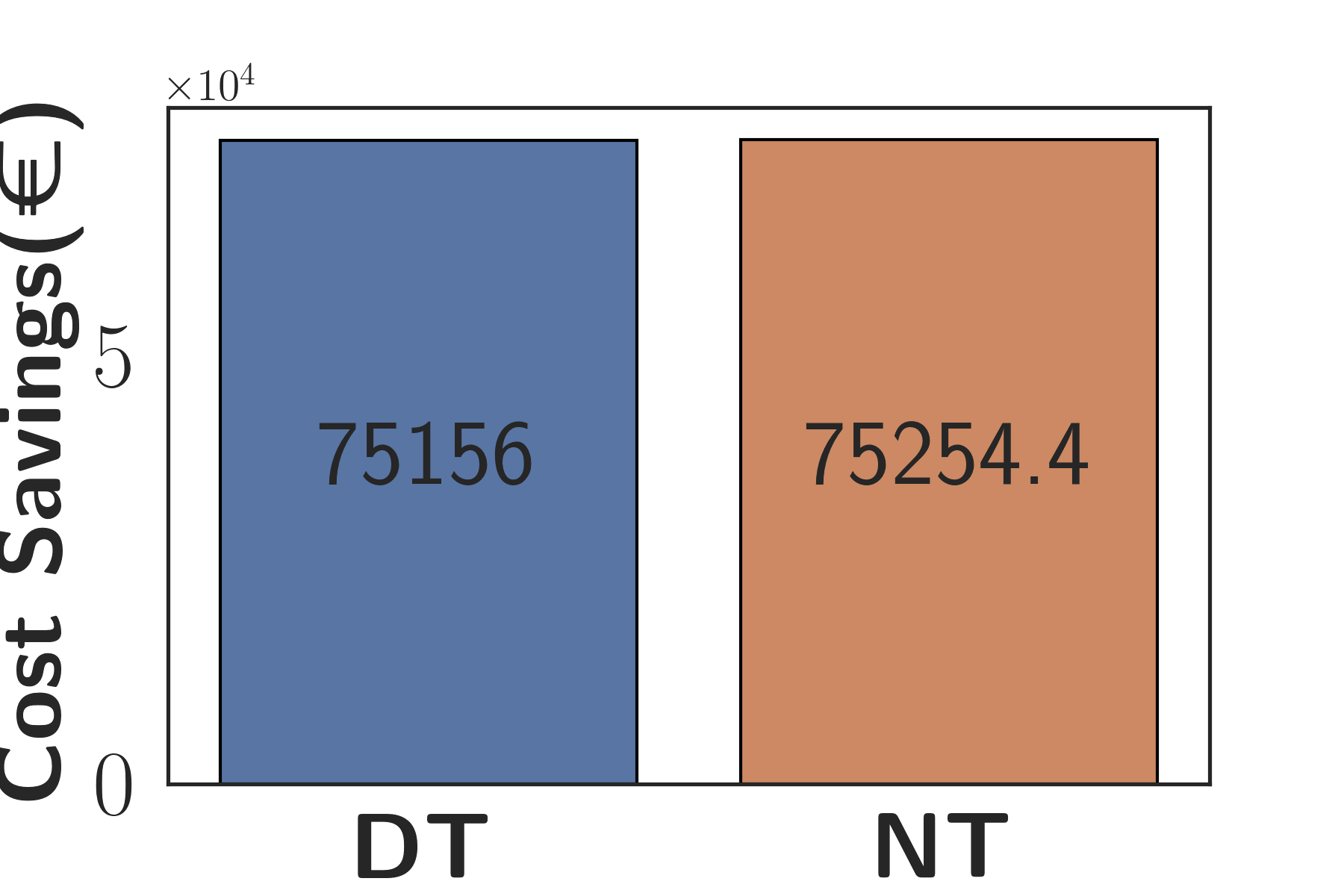}
\subcaption{ Pre-training vs Direct }
\label{fig:near_transfer}
	\end{minipage}
		\caption[MLOps Workflows]{MLOps workflows.}
  \label{fig:additional_mlops_workflows}
\end{figure}

We now present two specific workflows:
\textbf{Exploit  domain knowledge or other agents to learn better policies: }
Energy operators have significant historical data in terms of decision-making either manually or using any traditional techniques. In cases where the data available to train the RL agent is sparse, one can use this historical decision dataset along with the new incoming data to train an offline RL agent. Offline-RL in principle leverages large, previously collected datasets to learn effective policies without the need to constantly interact with the environment as required by traditional RL.

To show this, we first use MILP to solve the objective functions of the scenario by maintaining the constraints on the historically available datasets. For example, here, we run MILP for the first 2 years of UK data and then, make our own Markov Decision Process-based dataset (MDPDataset) for the data-driven offline RL approach which is used like a supervised learning dataset. We leverage \textit{conservative Q-learning} (CQL) provided by \textit{d3rlpy}, a python-based offline RL library. Figure~\ref{fig:offline-rl} shows that using Offline-RL the savings increases by 11\% (83K to 92K EUR) as compared to training an RL agent from scratch (DQN-RL).

\textbf{Use pre-trained models for new configurations:}
We now show how a pre-trained model with a specific action space can be re-used with a scenario with different action spaces. Specifically, we take two battery configurations, battery-1: 3 action spaces (charge at max, discharge at max or stay idle) and battery-2: 5 action spaces, in addition to the above three we have two new actions, i.e., (charge at half rate and discharge at half rate). Since the task is the same we finetune the pre-trained model with a 3-state battery configuration to target the task with a 5-state configuration. Figure~\ref{fig:near_transfer} shows that with pre-training and near-transfer we can achieve significant savings as opposed to direct training.

\end{document}